\documentclass[pra,nofootinbib, twocolumn,floatfix,superscriptaddress]{revtex4-2}  
\expandafter\let\csname equation*\endcsname\relax
\expandafter\let\csname endequation*\endcsname\relax

\usepackage{graphicx}  
\usepackage{orcidlink} 
\usepackage{amsfonts, amssymb}   
\usepackage{bm}        
\usepackage{amsmath}    
\usepackage{amsthm}    
\usepackage{verbatim}   
\usepackage{xcolor}      
\usepackage{hyperref}   
\usepackage[caption=false]{subfig}

\usepackage{float}
\usepackage{physics}
\usepackage{enumerate}
\usepackage{mathrsfs}
\usepackage{relsize}
\usepackage{mathtools}
\usepackage{wasysym}
\usepackage{tikz}
\usetikzlibrary{graphs, arrows,chains,matrix,positioning,scopes,shadows}
\usepackage{scalerel,stackengine}
\usepackage{accents}
\usepackage{natbib}

\stackMath
\newcommand{\stretchedhat}[1]{%
\savestack{\tmpbox}{\stretchto{%
  \scaleto{%
    \scalerel*[\widthof{\ensuremath{#1}}]{\kern.1pt\mathchar"0362\kern.1pt}%
    {\rule{0ex}{\textheight}}
  }{\textheight}%
}{2.4ex}}%
\stackon[-6.9pt]{#1}{\tmpbox}%
}
\newcommand{\stretchedtilde}[1]{%
\savestack{\tmpbox}{\stretchto{%
  \scaleto{%
    \scalerel*[\widthof{\ensuremath{#1}}]{\kern.1pt\mathchar"307E\kern.1pt}%
    {\rule{0ex}{\textheight}}
  }{\textheight}%
}{2.4ex}}%
\stackon[-6.9pt]{#1}{\tmpbox}%
}

\theoremstyle{definition}

\newcommand{\half}{\frac{1}{2}}

\newcommand{\imi}{\mathrm{i}}

\newcommand{\PTr}[2]{\text{tr}_{#1}\left(#2\right)}
\newcommand{\Hil}{{\mathcal H}}
\newcommand{\Th}{\text{th}}
\newcommand{\J}{{J\vphantom{\overline{J}}}}
\newcommand{\Jbar}{{\overline{J}}}
\renewcommand{\d}{{\text{d}}}
\newcommand{\Boltz}{ k_{\rm\scriptscriptstyle B}}
\definecolor{rkrPurple}{HTML}{73024F}

\begin{document}

\title{Dissipation in fermionic two-body continuous-time quantum walk under the steepest entropy ascent formalism}
\author{Rohit Kishan Ray\orcidlink{0000-0002-5443-4782}}
\email{rkray@ibs.re.kr}
\affiliation{Center for Theoretical Physics of Complex Systems, Institute for Basic Science (PCS-IBS), Daejeon - 34126, South Korea}

\affiliation{Theoretical Science Division, Poornaprajna Institute of Scientific Research, Bengaluru, Karnataka - 562110, India}
\author{R. Srikanth\orcidlink{0000-0001-7581-2546}}
\email{srik@ppisr.res.in}
\affiliation{Theoretical Science Division, Poornaprajna Institute of Scientific Research, Bengaluru, Karnataka - 562110, India}
\author{Sonjoy Majumder\orcidlink{0000-0001-9131-4520}}
\email{sonjoym@phy.iitkgp.ac.in}
\affiliation{Department of Physics, Indian Institute of Technology Kharagpur, Kharagpur, West Bengal - 721302, India}

\date{\today}


\begin{abstract}
  Quantum  walks play a crucial role in quantum algorithms and computational problems. Many-body quantum walks can reveal and exploit quantum correlations that are unavailable for single-walker cases. Studying quantum walks under noise and dissipation, particularly in multi-walker systems, has significant implications. In this context, we use a thermodynamically consistent formalism of dissipation modeling, namely the steepest entropy ascent (SEA) formalism. We analyze two spinless fermionic continuous-time walkers on a 1D graph with tunable Hubbard and extended Hubbard-like interactions. By contrasting SEA-driven dynamics with unitary evolution, we systematically investigate how interaction strengths modulate thermalization and entropy production. Our findings highlight the relevance of SEA formalism in modeling nonlinear dissipation in many-body quantum systems and its implications for quantum thermalization.
\end{abstract}

\maketitle
\section{\label{sec:intro}Introduction}
Recent progress in quantum information processing, quantum algorithms, quantum protocols, and their applications can be manifested using different quantum walk models. First introduced by \citet{aharonov_1993_quantum} in 1993, and later utilized as a search tool by \citet{shenvi_2003_quantum} and \citet{childs_2004_spatial}, the quantum walk algorithm has come a long way since then. \citet{childs_2009_universala} showed that quantum walks present a universal model of quantum computation. Quantum walks are involved in the modeling of relativistic dynamics \cite{chandrashekar2010relationship} as well as thermalization, including understanding eigenvalue thermalization \cite{dhamapurkar_2023_quantum}. \citet{duda_2023_quantum} studied diffusion, and localization on random lattices using quantum walks. Quantum walks can route entanglement on a network \cite{gao_2023_demonstration}, and find application in quantum magnetometry \cite{shukla2024quantum}.\\
Many-body physics can be explored via multi-walker quantum walks (MWQW). One of the first studies in MWQW was the two-walker (either entangled or otherwise) walks on a line  \cite{omar_2006_quantum,venegas-andraca_2009_quantum}. \citet{childs_2013_universal} showed that MWQW is also a universal model of quantum computation. \citet{rohde_2011_multiwalkerb} did a detailed study of the multi-walker formalism on the graphs and their photonic implementation. \citet{xue_2012_two} showed that sharing a coin between two walkers increases mutual information as swapping increases. MWQW is being used to model flexible teleportation schemes for multi-qubit systems \cite{li_2019_new}. Quantum foundation problems, such as the study of non-locality and local realism models, have been investigated using multi-walker quantum walks \cite{orthey_2019_nonlocality}. Recently, quantum walks have been implemented on IBM quantum computer \cite{acasiete_2020_implementation}. \citet{jiao_2021_twodimensional} showed that MWQW with photons on a two-dimensional lattice is used to simulate various genuine quantum phenomena. An MWQW has been used to mimic the effects of gravitationally induced entanglement \cite{badhani_2021_gravitationally}. Recently, \cite{dey_2023_collaboration} has analyzed collaborative quantum walks with more than two walkers. Two-walker quantum walks have been used for the quantum color image encryption protocol \cite{su_2023_quantum}.\\
Dissipation in quantum walks can result from experimental noise or environmental interactions, altering system behavior. Dissipative studies follow two main approaches. The first is the widely used Lindbladian formalism, which ensures complete positivity and trace preservation \cite{gorini_1976_completely, kumar2018non}. Here, the system weakly couples to the environment while the combined system evolves unitarily. Under the Markovian assumption, the system-environment state starts as a product state, leading to an irreversible yet thermodynamically consistent evolution based on specific environmental models---a `bottom-up’ approach \cite{ray_2025_nosignalinga}. Most dissipative quantum walk studies use this approach via the Gorini-Kossakowski-Lindblad-Sudarshan (GKLS) master equation \cite{kosloff_2019_quantum, chruscinski_2022_dynamical}. \citet{kendon_2003_decoherence} first explored decoherence in quantum walks under this framework (see \cite{kendon_2007_decoherence, fedichkin_2021_analysis} for reviews). \citet{fedichkin_2005_mixing} analyzed decoherence via mixing, while \citet{candeloro_2020_continuoustime} studied continuous-time quantum walks (CTQW) under quadratic Hamiltonian perturbations. \citet{garnerone_2012_thermodynamic} investigated thermodynamic properties, and \citet{pegoraro_2022_effects} recently examined conditioned losses in two-photon walks.\\
We propose to adopt a `top-down’ approach, starting with model dynamics in the density operator formalism to derive a thermodynamically consistent master equation. This approach derives a general, thermodynamically consistent master equation where the Gibbs state is the globally stable equilibrium (per the second law of thermodynamics). This leads to a nonlinear dynamical equation without exotic effects like signaling \cite{ray_2025_nosignalinga}. A key candidate for this approach is the steepest entropy ascent (SEA) formalism, proposed by \citet{beretta_1984_quantuma} to unify thermodynamics and mechanics \cite{hatsopoulos_1976_unified,*hatsopoulos_1976_unifieda,*hatsopoulos_1976_unifiedb,*hatsopoulos_1976_unifiedc}. SEA evolution for composite systems was later introduced \cite{beretta_1985_quantuma}, and Beretta demonstrated its thermodynamic consistency and applicability to general quantum dissipation \cite{beretta_2005_nonlinear,beretta_2009_nonlinear,beretta_2010_maximuma,beretta_2006_nonlineara}. He also proposed a generalized SEA framework similar to other dissipative models \cite{beretta_2014_steepesta}, eventually arguing that SEA could be considered the fourth law of thermodynamics \cite{beretta_2020_fourth}. Beyond pedagogical advances, SEA has seen growing applications. It has been used for temperature and magnetization modeling in low-temperature systems \cite{yamada_2019_lowtemperature} and for predicting entanglement loss in controlled phase gates \cite{montanez-barrera_2020_lossofentanglement}. One of the authors applied SEA to study dissipation in CTQW and developed an approximate analytical method using fixed Lagrange multipliers (FLM) \cite{ray_2022_steepestb}. SEA has also been used to model decoherence in superconducting quantum processors \cite{montanez-barrera_2022_decoherence}, dissipative dynamics in two-qubit gates \cite{tabakin_2023_local}, and non-local correlation loss, with \citet{damian_2024_modeling} showing strong agreement between SEA predictions and experiments. Despite its nonlinearity, SEA evolution does not lead to signaling \cite{ray_2025_nosignalinga}.\\
From the discussion of the preceding paragraphs we conclude that the important problem of dissipation in MWQW has not been explored through the SEA framework. In this paper, we study the problem of dissipation in two-walker CTQW under the SEA evolution. One of the major advantages of using SEA lies in the fact that one does not need to worry about particular modeling of the environment --- the relaxation dynamics will continue to drive the system towards the available maximal entropic state via a path of steepest entropy production. We study two fermionic particles walking on a ring (1D lattice with periodic boundaries) and analyze the evolution of the dissipative walk. Additionally, we examine the effect of SEA on MWQW across different interaction regimes (Table \ref{table:cases}), considering Hubbard and extended Hubbard-like interactions with varying strengths.\\
This paper is organized as follows. In Sec. \ref{sec:theory}, we introduce the necessary theoretical background for this work. We introduce the continuous-time quantum walker for two-walker in section \ref{secsub:CTQW}, and in section \ref{secsub:SEA} we do the same for the steepest entropy ascent formalism. In section \ref{sec:two_walker_sea}, we discuss the various regimes of interaction under consideration and present the results of our analysis. We discuss the results in section \ref{sec:discussions} and therein present our concluding remarks.


\section{\label{sec:theory}Theoretical preliminaries}
\subsection{\label{secsub:CTQW}The two-walker continuous-time quantum walk}
\begin{figure}[t!]
\centering
\includegraphics[width = \columnwidth]{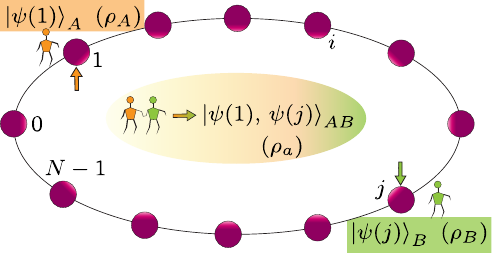}
\caption{\label{fig:two_walker} A schematic of the two-walker model on a ring graph of $N$ vertices indexed from $0$ to $N-1$. The two-walker wave-function is an element of the joint Hilbert space $\mathcal{H} = \mathcal{H}_A \otimes \mathcal{H}_B$. Initially, walkers $A$ and $B$ are localized in distinct regions, allowing the composite state to be written as a product state. As correlations develop during evolution, the system is described by a composite antisymmetric density matrix ($\rho_a$). The reduced density matrix $\rho_\J$ represents the \(\J^\Th\) walker.}
\end{figure}

The single quantum walker can be modeled on an underlying graph. We begin by considering an undirected graph $\mathcal{G}$ with no loops and multiple edges. $\mathcal{G}$ has a vertex set $\mathbb{V}$ with $N$ vertices and an edge set $\mathbb{E}$ defined as the set of edges that exist between the vertices. We associate a degree matrix $\bm{D}$, a diagonal matrix with $i^\Th$ entry denoting the degree (number of edges incident on a given vertex) of the $i^\Th$ vertex (see Fig. \ref{fig:two_walker}). Using an adjacency matrix $\bm{A}$ defined as follows
\begin{equation}\label{eq:adjacency}
  \bm{A}: a_{ij} = \begin{cases}
    1 & \text{if } e_{ij}\in \mathbb{E} \\
    0 & \text{otherwise}
  \end{cases},
\end{equation}
we define the Laplacian $\bm{L}$ on $\mathcal{G}$ using the relation: $\bm{L} = \bm{D}-\bm{A}$. We can write the following equation of motion for a continuous-time quantum walker\cite{ray_2022_steepestb},
\begin{equation}\label{eq:unitary_one_walker}
  \dv{\rho}{t} = -\dfrac{\imi}{\hbar}\comm{H} {\rho} = -\dfrac{\imi}{\hbar}\comm{\bm{\mu L}}{\rho}.
\end{equation}
Using the computational basis, we express $\rho = \sum_i p_i \dyad{i}{i}$, where $p_i$ is the probability that the walker is at $v_i$. We notice that the effective Hamiltonian describing the evolution can be written as $\mathit{H} = \bm{\mu L}$, $\bm{\mu}$ a square matrix of size $N$ that contains the hopping-probability and on-site potential information. We can express the same Hamiltonian in tight-binding form as 
\begin{align}\label{eq:tb_graph}
  \begin{split}
    \bm{L} &= \sum_{i=0}^{N-1}d_i\dyad{i}{i}- \sum_{\langle i,j \rangle}\left(\dyad{i}{j}+\dyad{j}{i}\right),\\
    \Rightarrow \mathit{H} &= \sum_{i=0}^{N-1}d_i\mu_{ii}\dyad{i}{i}- \sum_{\langle i,j \rangle}\mu_{ij}\left(\dyad{i}{j}+\dyad{j}{i}\right),\\
     &= \sum_{i=0}^{N-1}\epsilon_i\dyad{i}{i}- t\sum_{\langle i,j \rangle}\left(\dyad{i}{j}+\dyad{j}{i}\right).
  \end{split}
\end{align}
We have used hopping probability $\mu_{ij}$ to denote the transition probability per unit time between two vertices $v_i$ and $v_j$ (with an additional assumption of uniform transition probability, $\mu_{ij}=\mu$), $d_i=2$ is the degree of vertex $v_i$ and $\mu_{ii}=\mu$ for all $i$ for our purposes. Additionally, in our case of the ring graph, on-site potential term $\epsilon_i = d_i=2$ for all $i$. For simplicity, the nearest neighbor hopping term $t$  is considered equal to $\mu =1$, for all pairs $\langle i,j \rangle$.
The solution to the equation Eq. (\ref{eq:unitary_one_walker}) is given by
\begin{equation}\label{eq:sol_uni_one_walk}
  \rho(t) \equiv \rho^t = \mathcal{U}_t\rho^0\mathcal{U}_t^\dagger,
\end{equation}
with $\mathcal{U}_t = \exp(-\imi \mathit{H}t)$ (in this paper, we consider $\hbar=1$).\\
We now do the straightforward extension of the above formalism to include the two-walker quantum walk. We consider walkers $A$ and $B$ walking on the same graph $\mathcal{G}$, the walk will be governed by a general Hamiltonian which includes a non-interacting and an interacting part in the following fashion.
\begin{equation}\label{eq:hamiltonian}
  \mathit{H} = \mathit{H}_\text{free}+\mathit{H}_\text{int},
\end{equation}
where $\mathit{H}_\text{free} = \mathit{H}_A\otimes\mathit{I}_B+\mathit{I}_A\otimes\mathit{H}_B$, and $\mathit{H}_\J, \mathit{I}_\J$ act on subsystem $\J$ for $\J \in A,B$. The term $\mathit{H}_\text{int}$ depends on the model of choice and will be discussed in Sec. \ref{sec:two_walker_sea}.
We consider two indistinguishable walkers with anti-symmetric wave-function defined in the computational basis as
\begin{equation}\label{eq:anti-wf}
  \ket{\psi_{ij}} = \dfrac{1}{\sqrt{2}}\left(\ket{ij}-\ket{ji}\right),
\end{equation}
and the corresponding density matrix as $\rho_{ij}=\dyad{\psi_{ij}}$.
We can also define projectors that will project onto the anti-symmetric subspace of the tensor-product Hilbert space $\mathcal{H}_A\otimes\mathcal{H}_B$. These projectors can be written in terms of swap operators $\mathit{S}\rho_{ij} = \rho_{ji}$, as
\begin{equation}\label{eq:projection}
  \mathrm{P}_a = \half \left(\mathit{I}-\mathit{S}\right).
\end{equation}
Using these projectors, the usual Scr\"{o}dinger-von Neumann equation of motion can be written as
\begin{equation}\label{eq:unitary_svn}
  \mathrm{P}_a\dv{\rho}{t}\mathrm{P}_a = -\imi \commutator{\mathit{H}_a}{\rho_a},
\end{equation}
where, $\mathit{H}_a = \mathrm{P}_a\mathit{H}\mathrm{P}_a$, and $\rho_a = \mathrm{P}_a\rho\mathrm{P}_a$.
Under this scheme, the unitary equation of motion of the density matrix can be written as (analogous to Eq. (\ref{eq:unitary_one_walker}))
\begin{equation}\label{eq:unitary_two_walker}
  \dv{\rho_a}{t} = -\imi\comm{\mathit{H}_a}{\rho_a}.
\end{equation}
This equation provides a guarantee that the system will be constrained to the relevant antisymmetric subspace throughout its evolution. The solution can be similarly written as,
\begin{equation}\label{eq:unitary_sol_two_W}
  \rho^t_a = \mathcal{U}_t\rho_a^0\mathcal{U}^\dagger_t,
\end{equation}
with $\mathcal{U}_t = \exp(-\imi\mathit{H}_at)$. The joint probability distribution (JPD) of the walkers at time $t$ is given by this $\mathcal{P}^t_a(m,n)$, and can be found as (for simultaneous detection at vertices or sites $m$ and $n$)
\begin{equation}\label{eq:jpd_two}
  \mathcal{P}^t_a(m,n) = \bra{mn}_{AB}\rho^t_a\ket{mn}_{AB}.
\end{equation}
The marginal probability of finding each of the $\J^\Th$ walker at time $t$ and at site $m$ can be given by ($\Jbar$ denotes the complementary system to $\J$),
\begin{equation}\label{eq:marginal_prob}
  p^t_a(m) = \bra{m}_\J\PTr{\Jbar}{\rho_a^t}\ket{m}_\J.
\end{equation}
We have used $\PTr{\Jbar}{\rho}$ to denote partial trace over the subsystem $\Jbar$.

\subsection{\label{secsub:SEA}Steepest entropy ascent formalism}
\subsubsection{\label{ssecsub:soft_intro} Single-component equation}
We present the theoretical background required for the steepest entropy ascent (SEA) formalism. However, for a detailed derivation and motivation for the SEA formalism, we direct the reader to the Refs. \cite{beretta_2014_steepesta,ray_2022_steepestb}. In the SEA formalism, the local entropy production is maximized in tandem with various conservation criteria. We begin by elaborating on the usage of the `top-down' term in the introduction.\\
The SEA dynamics describes the relaxation of a system from far-off equilibrium towards equilibrium. The Gibbs state is the stable equilibrium state from the canonical second law of thermodynamics \cite{hatsopoulos_1976_unified,*hatsopoulos_1976_unifieda,*hatsopoulos_1976_unifiedb,*hatsopoulos_1976_unifiedc}. A general dynamics that maximizes entropy production to reach such an equilibrium is essentially nonlinear \cite{simmons_1981_essential,beretta_1986_theorem}. This scheme also stands out as it considers a seldom confronted thermodynamic consistency criterion, the stability of the Gibbs state. Naturally, SEA becomes a `top-down' approach, as it does not build up from a Schr\"{o}dinger equation and derives the equation of motion (EoM) using the desiderata as following. Beretta formulated the original version of SEA in Refs. \cite{beretta_1984_quantuma,beretta_1985_quantuma,beretta_2010_maximuma,beretta_2009_nonlinear,beretta_2014_steepesta}, and hence we will call it the Beretta SEA (BSEA) EoM. In Ref. \cite{ray_2022_steepestb}, one of the authors has derived the BSEA EoM (see Appendices A and B therein). We will begin with the generic feature of the BSEA EoM, in the Ginzburg-Landau form \cite{li_2016_steepestentropyascent}:
\begin{equation}\label{eq:sea_gen_form}
  \dv{\rho}{t} = -\imi\comm{\mathit{H}}{\rho}-\acomm{\mathcal{D}}{\rho},
\end{equation}
where we have introduced the dissipation operator $\mathcal{D}$ in the anti-commutator of the RHS. In the absence of an external reservoir, the isolated system evolves in the direction of maximum local entropy production. As a consequence, the state vector evolves non-unitarily by strictly adhering to the constraints of the motion, while the trajectory moves more and more towards the global stable equilibrium state of the given context. Eq. (\ref{eq:sea_gen_form}) is a compactified form of BSEA, the full expression for $\mathcal{D}$ is given below\cite{beretta_2014_steepesta,ray_2022_steepestb}:
\begin{align}\label{eq:seacompact}
  \begin{split}
                                                           & \dv{\rho}{t} +\imi\comm{\mathit{H}}{\rho} =                                                                                                          \\
                                                           & -\dfrac{1}{\tau}\tfrac{\vmqty{\rho B\ln(\rho)             & \half\acomm{\mathit{C}_1}{\rho}                        & \half\acomm{\mathit{C}_2}{\rho} \\
    \tr(\tfrac{\rho}{2}\acomm{\mathit{C}_1}{B\ln(\rho)})   & \tr(\rho\mathit{C}_1^2)                                   & \tr(\tfrac{\rho}{2}\acomm{\mathit{C}_1}{\mathit{C}_2})                                   \\ \tr(\tfrac{\rho}{2}\acomm{\mathit{C}_2}{B\ln(\rho)}) &\tr(\tfrac{\rho}{2}\acomm{\mathit{C}_2}{\mathit{C}_1}) & \tr(\rho\mathit{C}_2^2)}}{\vmqty{\tr(\tfrac{\rho}{2}\acomm{\mathit{C}_1}{\mathit{C}_1}) &  \tr(\tfrac{\rho}{2}\acomm{\mathit{C}_1}{\mathit{C}_2})\\
    \tr(\tfrac{\rho}{2}\acomm{\mathit{C}_2}{\mathit{C}_1}) & \tr(\tfrac{\rho}{2}\acomm{\mathit{C}_2}{\mathit{C}_2})}}.
  \end{split}
\end{align}
In the above equation, $\tau$ is known as relaxation time, $\mathit{C}_i$ is the operator associated with the system conservation and constraint \textit{e.g.}, for a single particle $\mathit{C}_1$ is $\mathbf{I}$ operator, for probability conservation; and $\mathit{C}_2$ is the Hamiltonian operator $\mathit{H}$, for energy conservation. $\mathit{B}$ is an idempotent operator, projecting $\ln(\rho)$ on the kernel of $\rho$, making it analytically well defined. $\mathit{B}$ can be formally written as $\mathit{B}=P_{\ker(\rho)}$. We cast Eq. (\ref{eq:seacompact}) in the following convenient form,
\begin{equation}\label{eq:seaconvenient}
  \dv{\rho}{t} = -\imi\comm{\mathit{H}}{\rho} -\dfrac{1}{2\tau}\left[\acomm{B\ln(\rho)}{\rho}+\sum_i(-1)^i\beta_i\acomm{\mathit{C}_i}{\rho}\right].
\end{equation}
Here, the parameters, $\beta_i$, are defined explicitly in Eq. (\ref{eq:seacompact}) and, in general, are nonlinear functionals of $\rho$, which vary in time during the evolution. In the presence of reservoir, $\beta_2$ associated with $\mathit{H}$ can be interpreted as inverse temperature and is solely determined by the reservoir \cite{damian_2024_modeling}. In some cases, as discussed in \cite{ray_2022_steepestb}, these $\beta_i$'s can be considered constant and that consideration reduces the nonlinearity present in Eq. (\ref{eq:seaconvenient}), especially in the low $\tau$ region. We define the operator $\mathcal{D}$,
\begin{align}\label{eq:wseaop}
  \mathcal{D} = & \dfrac{1}{2\tau}\left[\mathit{B}\ln(\rho)+\sum_i(-1)^i\beta_i\mathit{C}_i\right].
\end{align}
We use $\mathcal{D}$ to write equation (\ref{eq:seaconvenient}) as Eq. (\ref{eq:sea_gen_form}). This completes the short introduction of the single-component BSEA EoM. However, for our purposes, we need to use the two-component BSEA EoM.

\subsubsection{\label{ssecsub:hard_intro}The two-component equation}
Before proceeding with the two-component BSEA EoM, we need to address the subtleties of using a non-linear evolution to describe many-body dynamics. In interacting systems, the interaction energy, and in correlated systems, the mutual entropy (as defined later), do not have a clear division between system components \cite{beretta_1985_quantuma,beretta_2010_maximuma}. Meanwhile, SEA dynamics maximizes local entropy production. Without a proper framework to define the local contribution of these quantities, implementing SEA evolution becomes challenging. In this regard, we use the `local-perception' operators (LPOs) \cite{beretta_1985_quantuma,beretta_2009_nonlinear,beretta_2010_maximuma,ray_2025_nosignalinga} for the following reasons:
\begin{enumerate}
  \item Unlike the linear Scr\"{o}dinger-von Neumann formalism, which retains the same form of EoM for both composite and single systems, the BSEA EoM, being nonlinear, needs to respect the structure of the composite to avoid unphysical interactions \cite{ray_2025_nosignalinga}.
  \item The LPOs, constructed via a weighted projection of the composite operator onto local Hilbert spaces (Eq. (\ref{eq:loc_perc_ci})), are no-signaling, as their expectation values remain unchanged under local unitary operations in other subsystems \cite{ray_2025_nosignalinga}.
\end{enumerate}
We consider the dynamical equation of composite systems in the following manner. Consider the Hilbert space of the $N$ partite composite system of the form $\mathcal{H}_1\otimes\mathcal{H}_2\otimes\cdots\otimes\mathcal{H}_N$. The SEA formalism is built on the `locally' steepest entropy ascent, maximizing the locally `perceived' entropy and conserving corresponding `perceived' constraint functionals \cite{beretta_1985_quantuma,beretta_2010_maximuma,ray_2025_nosignalinga}. As a result, each of these local subsystems undergoes SEA treatment. The general Hamiltonian has the form $\mathit{H} = \sum_\J\mathit{H}_\J\otimes\mathit{I}_\Jbar+\mathit{V}$, where $\mathit{V}$ is the interaction term, $\mathit{H}_\J$ the local Hamiltonian of the $\J^\Th$ subsystem in $\Hil_\J$, and $\mathit{I}_\Jbar \in \Hil_\Jbar = \bigotimes_{K\ne \J}\Hil_K$. The reduced density matrices of the $\J^\Th$ component is $\rho_\J = \PTr{\Jbar}{\rho}$. The LPO, as originally introduced in Ref. \cite{beretta_1985_quantuma} and recently reintroduced in the context of no-signaling and quantum information tasks in Ref. \cite{ray_2025_nosignalinga} is defined as -
\begin{equation}\label{eq:loc_perc_ci}
  (\mathit{C}_i)^\J = \PTr{\Jbar}{(\mathit{I}_\J\otimes\rho_{\Jbar})\mathit{C}_i}.
\end{equation}
We immediately notice that for a two-component system, $AB$, the LPOs defined on subsystems $A$ and $B$ are unique and express the limitation of the information $A$ and $B$ can have about the overall operator $\mathit{X}$ via classical communication. This can be expressed as
\begin{equation}\label{eq:local_perception_operators_id}
  \Tr[\rho_A(X)^A_\rho] =\Tr[(\rho_A\otimes\rho_B)X] =\Tr[\rho_B(X)^B_\rho].
\end{equation}
We define the locally perceived entropy operator (LPEO) as
\begin{align}\label{eq:loc_perc_S}
  \begin{split} (\mathit{S}(\rho))^A & = \PTr{B}{(\mathrm{I}_2\otimes\rho_B)\mathit{S}(\rho)}                                                                                                      \\
              (\mathit{S}(\rho))^B & = \PTr{A}{(\rho_A\otimes\mathrm{I}_2)\mathit{S}(\rho)}                                                                                                                    \\
              \mathit{S}(\rho)     & = -\Boltz\mathrm{B}\ln(\rho) \mbox{ with } \mathrm{B}\ln(x) = \left\{\begin{matrix} 0 &\mbox{ for } x\le 0\\ \ln(x) &\mbox{ for } x> 0 \end{matrix} \right.
  \end{split}
\end{align}
We impose another salient feature of the evolution to ensure SEA is trace-preserving (TP), the part in the anticommutator of RHS in Eq. (\ref{eq:sea_gen_form}) must be traceless (similar to Lindblad evolution). We can extend this to the case of a many-body SEA equation and demand that the local dissipative operators $\mathcal{D}_\J$ be such that $\acomm{\mathcal{D}_\J}{\rho_\J}$ is traceless. We can write the many-body BSEA EoM (for $M$ constituents) as  \cite{beretta_1985_quantuma,beretta_2005_nonlinear,beretta_2009_nonlinear,beretta_2010_maximuma,ray_2025_nosignalinga}
\begin{equation}\label{eq:comp_sea_simplified}
  \dv{\rho}{t} = -\imi\comm{\mathit{H}}{\rho}-\sum_{\J=1}^M\acomm{\mathcal{D}_\J}{\rho_\J}\otimes\rho_\Jbar.
\end{equation}
We note that $\mathcal{D}_\J$ operates on $\Hil_\J$, and is nonlinear. Also, if one sets $M=1$, we can recover Eq. (\ref{eq:sea_gen_form}). The expression for $\mathcal{D}_\J$ can be written as (from Appendix \ref{app:two_component}) \cite{beretta_1985_quantuma,ray_2025_nosignalinga},
\begin{widetext}
  \begin{equation}\label{eq:DJ_compact}
    \acomm{\mathcal{D}_J}{\rho_J} = \dfrac{1}{2\tau_J}\tfrac{\vmqty{\rho_J(B\ln(\rho))^J & \half\acomm{(\mathit{C}_1)_J}{\rho_J} &  \half\acomm{(\mathit{C}_2)_J}{\rho_J}\\
        \tr(\tfrac{\rho_J}{2}\acomm{(\mathit{C}_1)_J}{(B\ln(\rho))^J}) & \tr(\rho_J(\mathit{C}_1)_J^2) &  \tr(\tfrac{\rho_J}{2}\acomm{(\mathit{C}_1)_J}{(\mathit{C}_2)_J})\\ \tr(\tfrac{\rho_J}{2}\acomm{(\mathit{C}_2)_J}{(B\ln(\rho))^J}) &\tr(\tfrac{\rho_J}{2}\acomm{(\mathit{C}_2)_J}{(\mathit{C}_1)_J}) & \tr(\rho_J(\mathit{C}_2)_J^2)}}{\vmqty{\tr(\tfrac{\rho_J}{2}\acomm{(\mathit{C}_1)_J}{(\mathit{C}_1)_J}) &  \tr(\tfrac{\rho_J}{2}\acomm{(\mathit{C}_1)_J}{(\mathit{C}_2)_J})\\
        \tr(\tfrac{\rho_J}{2}\acomm{(\mathit{C}_2)_J}{(\mathit{C}_1)_J}) & \tr(\tfrac{\rho_J}{2}\acomm{(\mathit{C}_2)_J}{(\mathit{C}_2)_J})}}
  \end{equation}
\end{widetext}
We note that the local Lagrange multipliers can be computed as in (from Appendix \ref{app:two_component})
\begin{equation}\label{eq:app1_omegaJ}
  \Omega^J =  \vmqty{\tr(\tfrac{\rho_J}{2}\acomm{(\mathit{C}_1)_J}{(\mathit{C}_1)_J}) &  \tr(\tfrac{\rho_J}{2}\acomm{(\mathit{C}_1)_J}{(\mathit{C}_2)_J})\\
    \tr(\tfrac{\rho_J}{2}\acomm{(\mathit{C}_2)_J}{(\mathit{C}_1)_J}) & \tr(\tfrac{\rho_J}{2}\acomm{(\mathit{C}_2)_J}{(\mathit{C}_2)_J})},
\end{equation}
and then
\begin{align}\label{eq:app1_betaJ1}
  \begin{split}
                                                                   & \beta_1^J =                                                                                                                                                 \\
                                                                   & \dfrac{1}{\Omega^J}\vmqty{\tr(\tfrac{\rho_J}{2}\acomm{(\mathit{C}_1)_J}{(B\ln(\rho))^J}) & \tr(\tfrac{\rho_J}{2}\acomm{(\mathit{C}_1)_J}{(\mathit{C}_2)_J}) \\
    \tr(\tfrac{\rho_J}{2}\acomm{(\mathit{C}_2)_J}{(B\ln(\rho))^J}) & \tr(\tfrac{\rho_J}{2}\acomm{(\mathit{C}_2)_J}{(\mathit{C}_2)_J})},
  \end{split}
\end{align}
\begin{align}\label{eq:app1_betaJ2}
  \begin{split}
                                                                   & \beta_2^J=                                                                                                                                                  \\
                                                                   & \dfrac{1}{\Omega^J}\vmqty{\tr(\tfrac{\rho_J}{2}\acomm{(\mathit{C}_1)_J}{(B\ln(\rho))^J}) & \tr(\tfrac{\rho_J}{2}\acomm{(\mathit{C}_1)_J}{(\mathit{C}_1)_J}) \\
    \tr(\tfrac{\rho_J}{2}\acomm{(\mathit{C}_2)_J}{(B\ln(\rho))^J}) & \tr(\tfrac{\rho_J}{2}\acomm{(\mathit{C}_2)_J}{(\mathit{C}_1)_J})}.
  \end{split}
\end{align}
Thus we can write the following simplified expression for the local SEA dissipation operator,
\begin{equation}\label{eq:DJ_Def}
  \mathcal{D}_J = \dfrac{1}{2\tau_J}\left((\mathit{B}\ln(\rho))^J+\sum_i(-1)^i\beta^J_i(\mathit{C}_i)_J\right).
\end{equation}
Taking the partial trace over the Eq. (\ref{eq:comp_sea_simplified}) we get the model equation of dissipation for the subsystem $\J$ as \cite{beretta_2010_maximuma,ray_2025_nosignalinga}
\begin{equation}\label{eq:sea_gen_J}
  \dv{\rho_\J}{t} = -\imi\comm{\mathit{H}_\J}{\rho_\J} - \PTr{\Jbar}{\comm{\mathit{V}}{\rho}} - \acomm{\mathcal{D}_\J}{\rho_\J}.
\end{equation}
Now, as a final note, to account for the particle symmetries, we must include the projector just as we did in Eq. (\ref{eq:unitary_svn}). This implies projecting the overall Hilbert space and its operators to the subspace as required, and then compute the ``new'' local operators to implement the many-body BSEA EoM for the particular symmetry.
\begin{equation}\label{eq:symmetrized_BSEA}
  \dv{\rho_a}{t} = -\imi\comm{\mathit{H}_a}{\rho_a}-\sum_{\J=1}^{M}\acomm{\mathcal{D}_\J(\rho_a)}{(\rho_a)_\J}\otimes(\rho_a)_\Jbar
\end{equation}


\section{\label{sec:two_walker_sea}Two walkers under SEA}
\begin{table}[hb!]
  \centering
  \resizebox{\columnwidth}{!}{ 
  \begin{tabular}{|p{4cm}|p{2cm}|p{2cm}|p{2cm}|p{2cm}|} \hline
    \textbf{Case} & $\bm{\alpha_1}$ & $\bm{\alpha_2}$ & $\bm{\alpha_3}$ & $\bm{\alpha_4}$\\ 
    \hline
    \textbf{Full (FI)} (All equal) & $\neq 0$ & $\neq 0$ & $\neq 0$ & $\neq 0$\\
    \textbf{Hubbard (HI)} & $\neq 0$ & $=0$ & $=0$ & $=0$\\
    \textbf{Correlated Hopping}\newline \textbf{Interaction (CHI)} & $\neq 0$ & $=0$ & $=0$ & $\neq 0$\\
    \textbf{Full interaction with} \newline  \textbf{fixed hopping (FIFH)} & $\neq 0$ & $\neq 0$ and\newline $=\alpha_3$  & $\neq 0$ and\newline $=\alpha_2$ & $\neq0$\\
    \hline
  \end{tabular}
  }
  \caption{\label{table:cases}Major interaction regimes and corresponding conditions on $\alpha_i$'s in Eq. (\ref{eq:gen_total_ham}) for the two walkers.}
\end{table}
Before presenting our results, we elaborate on the structure of $\mathit{H}_{\text{int}}$. We start with a more general description, then allow modifications according to our requirements. A general form of interaction can be written down as 
\begin{equation}\label{eq:gen_int_comp}
  \mathit{H}_\text{(int, gen)} = \mathit{H}_A\otimes\mathit{H}_B,
\end{equation}
which upon expansion, using Hamiltonian as in Eq. (\ref{eq:tb_graph}), gives rise to the following expression (The indices $i,j$ belong to the walker $A$, while $k,\ell$ belong to the walker $B$. The hopping strengths $t, s$ and on-site potentials $\epsilon_i, \omega_k$ belong to $A$ and $B$, respectively.)
\begin{align}\label{eq:gen_int_full}
  \begin{split}
    \mathit{H}_\text{(int, gen)} &= \sum_{i,k}\epsilon_i\omega_k\dyad{ik}{ik}-t\sum_{\langle i,j\rangle, k}\omega_k\left(\dyad{ik}{jk}+\dyad{jk}{ik}\right)\\
    &-s\sum_{i,\langle k,\ell \rangle}\epsilon_i \left(\dyad{ik}{i\ell}+\dyad{i\ell}{ik}\right)\, \\
    & + ts\sum_{\langle i,j\rangle, \langle k,\ell \rangle}\left(\dyad{ik}{j\ell}+\dyad{i\ell}{jk}+\text{h.c.}\right). 
  \end{split}
\end{align}
We will also expand the term $\mathit{H}_\text{free}$ as given below
\begin{align}\label{eq:gen_free_full}
  \begin{split}
    \mathit{H}_\text{free} &= \mathit{H}_A\otimes\mathit{I}_B+\mathit{I}_A\otimes\mathit{H}_B,\\
    &= \sum_{i,k}\left(\epsilon_i+\omega_k\right)\dyad{ik}{ik}-t\sum_{\langle i,j\rangle, k}\left(\dyad{ik}{jk}+\dyad{jk}{ik}\right)\,\\
    &-s\sum_{i,\langle k,\ell \rangle}\left(\dyad{ik}{i\ell}+\dyad{i\ell}{ik}\right).
  \end{split}
\end{align}

Combining Eqs. (\ref{eq:gen_int_full}) and (\ref{eq:gen_free_full}), we get the total Hamiltonian as:
\begin{align}\label{eq:gen_total_ham}
  \begin{split}
    \mathit{H}_\text{total} &= \sum_{i,k}(\epsilon_i+\omega_k+\alpha_1\epsilon_i\omega_k)\dyad{ik}{ik}\,\\
    &-t\sum_{\langle i,j\rangle, k}(1+\alpha_2\omega_k)\left(\dyad{ik}{jk}+\dyad{jk}{ik}\right)\\
    &-s\sum_{i,\langle k,\ell \rangle}(1+\alpha_3\epsilon_i)\left(\dyad{ik}{i\ell}+\dyad{i\ell}{ik}\right)\,\\
    &+\alpha_4 ts\sum_{\langle i,j\rangle, \langle k,\ell \rangle}\left(\dyad{ik}{j\ell}+\dyad{i\ell}{jk}+\text{h.c.}\right).
  \end{split}
\end{align}
Where we introduce these scalars $\alpha_i$'s to tune each component of the above expression to include various levels of interaction.\\
We proceed by computing the projection of this Hamiltonian of Eq. (\ref{eq:gen_total_ham}) to the anti-symmetric subspace as we are interested in the fermionic type walkers. We get (using $\mathrm{P}_a$ as defined in Eq. (\ref{eq:projection}) ),

\begin{equation}\label{eq:ferm_tot_ham}
    \mathit{H}_\text{total, a}  = \mathrm{P}_a \mathit{H}_\text{total} \mathrm{P}_a
\end{equation}
\begin{figure}[t!]
  \centering
  \includegraphics[width=\columnwidth]{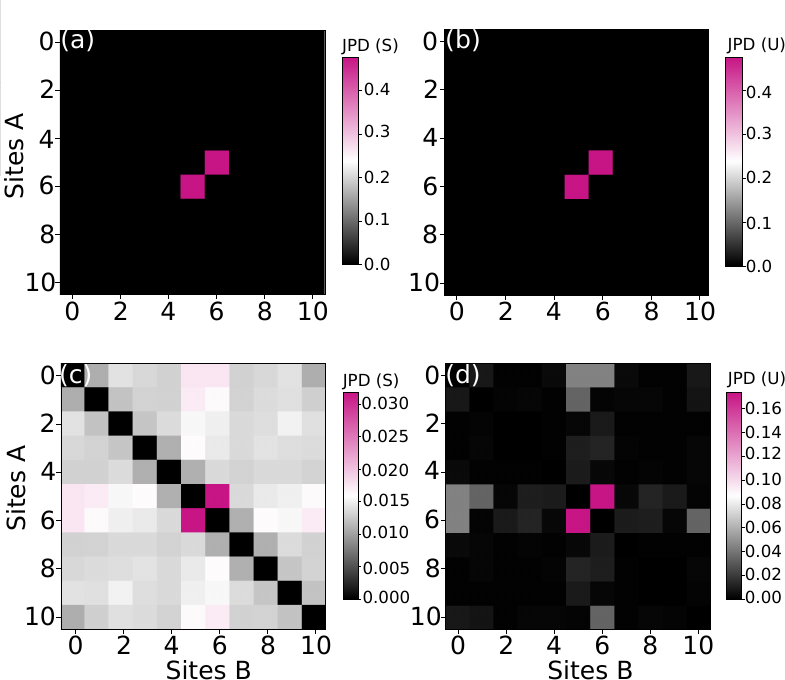}
  \caption{\label{fig:jpd_0_full}Joint probability distribution (JPD) of two-walker evolution on a ring with 11 nodes (indexed from zero). The walkers evolve without interaction ($\alpha_i=0$ $\forall$ $i$). Panels (a) and (b) show the initial JPD, while panels (c) and (d) depict the JPD at $t/\tau=30$. SEA evolution is shown in (a) and (c), and unitary evolution in (b) and (d). Color bars next to each panel indicate the corresponding probability values.}
\end{figure}
Based on the different settings for the parameters ($\alpha_i$'s), we identify four major cases for our study. We compare Eq. (\ref{eq:gen_int_full}) and Eq. (\ref{eq:gen_free_full}) and note that the terms with $\alpha_2$ and $\alpha_3$ are already present in $\mathit{H}_\text{free}$, which is there to modify the weight of conditional hopping terms in the Hamiltonian. On the other hand, the term associated with $\alpha_1$ modifies the on-site contribution, so the real `interaction' term is due to the $\alpha_4$. Based on these observations, we find four regimes of interaction. First case is where all the $\alpha_i$'s are nonzero and equal, we call it `full interaction'(FI). In the case where only $\alpha_1$ is non-zero, we call it the `Hubbard'(HI) regime (because of the spinless antisymmetric consideration, there are no true Hubbard interactions). For only non-zero values of $\alpha_1$ and $\alpha_4$, we get a `correlated hopping interaction'(CHI) regime. And finally we consider fixed conditional hopping terms ($\alpha_2=\alpha_3$) with $\alpha_1$ and $\alpha_4$ varying equally, we call this `full interaction with fixed hopping'(FIFH). The chosen values are within three orders of magnitudes, so weak interaction implies the value $0.1$, medium $1$, and strong at $10$. $t$ and $s$ are considered to be one. We summarize this classification in the table \ref{table:cases}.
\begin{figure}[t!]
  \centering
  \includegraphics[width=\columnwidth]{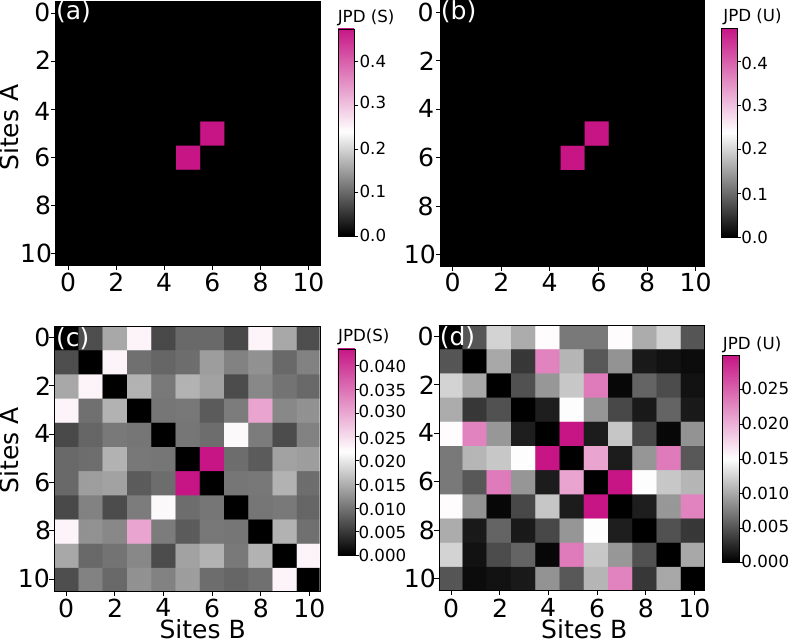}
  \caption{\label{fig:jpd_10_full}Joint probability distribution (JPD) of two-walker evolution (see Table \ref{table:cases}) on a ring with 11 nodes (indexed from zero). The walkers experience strong, full interaction ( $\alpha_i = 10$ $\forall$ $i$). Panels (a) and (b) show the initial JPD, while panels (c) and (d) depict the JPD at $t/\tau=30$. SEA evolution is shown in (a) and (c), and unitary evolution in (b) and (d). Color bars next to each panel indicate the corresponding probability values.}
\end{figure}
\begin{figure*}[t!]
  \centering
  \includegraphics[width=\textwidth, height=15cm]{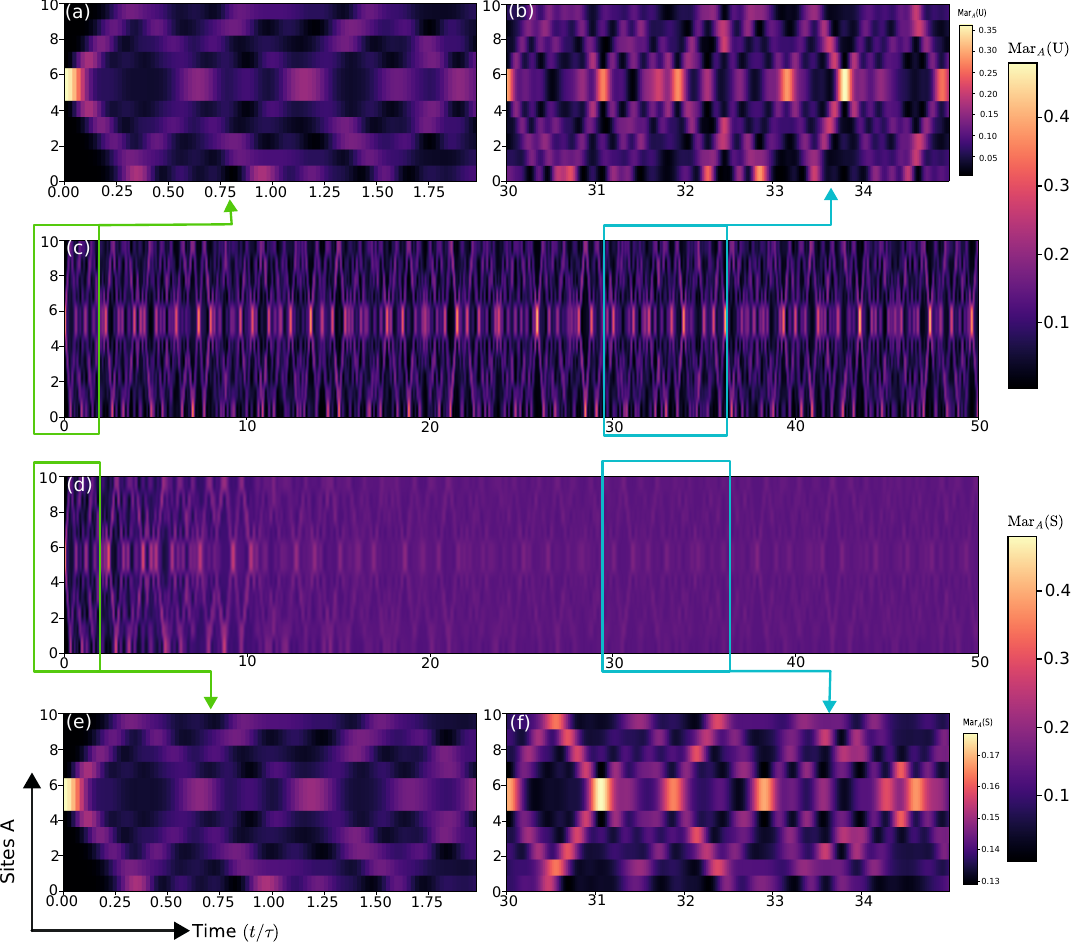}
  \caption{\label{fig:marg_0_full}Marginal probability of walker A on a ring graph with 11 sites, evolving under a Hamiltonian with no interaction. Panels (a)--(c) depict unitary evolution, while panels (d)--(f) show SEA evolution. Panels (a) and (e) compare the initial evolution of the marginal probability distribution under unitary and SEA dynamics, respectively. Panels (b) and (f) show the late-time evolution for the same cases. The large color bars on the left of each panel indicate probability values, while the smaller ones on the left of panels (b) and (f) correspond to zoomed-in probability scales. The x-axis represents time $(t/\tau)$, and the y-axis denotes the site number. Each of the smaller panels ((a), (b)) and ((e), (f)) are zoomed portions of panels (c) and (d), respectively, with the zoomed-in sections marked by rectangles.}
\end{figure*}
\begin{figure*}[t!]
  \centering
  \includegraphics[width=\textwidth, height=15cm]{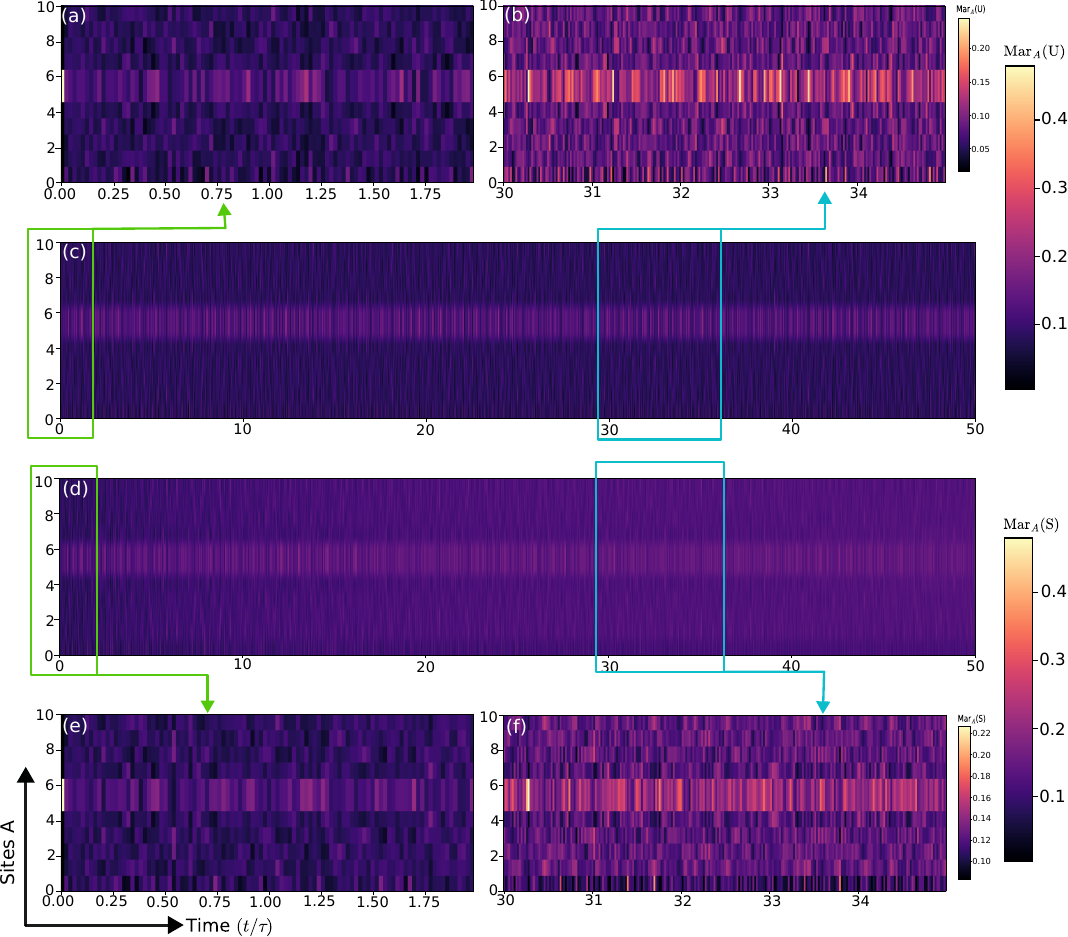}
  \caption{\label{fig:marg_10_full}Marginal probability of walker A on a ring graph with 11 sites, evolving under the Full Interaction Hamiltonian with $\alpha_i = 10$ for all $i$s (see Table \ref{table:cases}). Panels (a)--(c) correspond to unitary evolution, while panels (d)--(f) depict SEA evolution. Panels (a) and (e) show the initial evolution of the marginal probability distribution for unitary and SEA dynamics, respectively, while panels (b) and (f) illustrate the late-time evolution. The large color bars on the left of each panel represent probability values, and the smaller ones on the left of panels (b) and (f) show corresponding zoomed-in probability scales. The x-axis represents time $(t/\tau)$, and the y-axis denotes the site number. Each of the smaller panels ((a), (b)) and ((e), (f)) are zoomed portions of panels (c) and (d), respectively, with the zoomed-in sections marked by rectangles.}
\end{figure*}
We will now study the dynamics of the two walkers in each case. We study the evolution in unitless time $t/\tau$, where $\tau$ is the average relaxation time defined as $\tau = (\tau_A+\tau_B)/2$. We first begin by understanding the unitary walk features in the case of various degrees of interaction. Our initial states are perturbed entangled states. We begin by having a `singlet' configuration (for the initial position of each walker being at either $i^\Th$ or $j^\Th$ site of the ring) of the form 
\[
\ket{\psi(i,j)} = \dfrac{1}{\sqrt{2}}\left(\ket{i}_A\ket{j}_B - \ket{j}_A\ket{i}_B\right),
\]
which is then perturbed by an amount $\varepsilon \in [0,1)$ by a white noise (uniformly distributed over all the basis states spanning the antisymmetric subspace of $\mathcal{H}_{a}=\mathcal{H}_A\wedge\mathcal{H}_B$) to generate mixed state as under:
\begin{equation}\label{eq:initial_rho}
\rho^0 = \varepsilon\dyad{\psi(i,j)}{\psi(i,j)}+(1-\varepsilon)\mathrm{I}_a.
\end{equation}
We use $\varepsilon=0.95$ to produce slight perturbation in the initial state. This is by no means the only way to create mixed states; there exist other approaches also \cite{damian_2024_modeling}. However, studying the effects of such methods lies beyond the scope of this work.\\
A reliable measure of two-walker evolution can be obtained by tracking the evolution of the joint probability distribution (JPD). We compute JPD via the Eq. (\ref{eq:jpd_two}). In Fig. \ref{fig:jpd_0_full}, we show how the JPD of the two-walker evolves without interaction. We present two time slices to show the difference in evolution; one is at $t/\tau = 0$ (panels (a) and (b)), and the other is at $t/\tau = 30$ (panels (c) and (d)) in Fig. \ref{fig:jpd_0_full}. The JPD evolves from two sharp peaks at (5,6) and (6,5) at $t/\tau = 0$ and spreads more under the SEA dissipation, in contrast to continuously oscillating unitary evolution. The underlying unitary feature is not totally lost during the SEA evolution (see the comparison between panels (c) and (d) in Fig. \ref{fig:jpd_0_full}), just that it is more smeared. In Fig. \ref{fig:jpd_10_full}, where we are in the full interaction range (Table. \ref{table:cases}) with $\alpha_i = 10$ for all $i$s. We see that the SEA evolution (panel (c) of Fig. \ref{fig:jpd_10_full}) has lesser probability peaks when compared to the unitary one (panel (d) of Fig. \ref{fig:jpd_10_full}). This can be seen by counting the number of magenta blocks in the respective panels. \\
\begin{figure}[t!]
\centering
 \subfloat[ \label{fig:msd_FI} FI]{\includegraphics[width=\columnwidth]{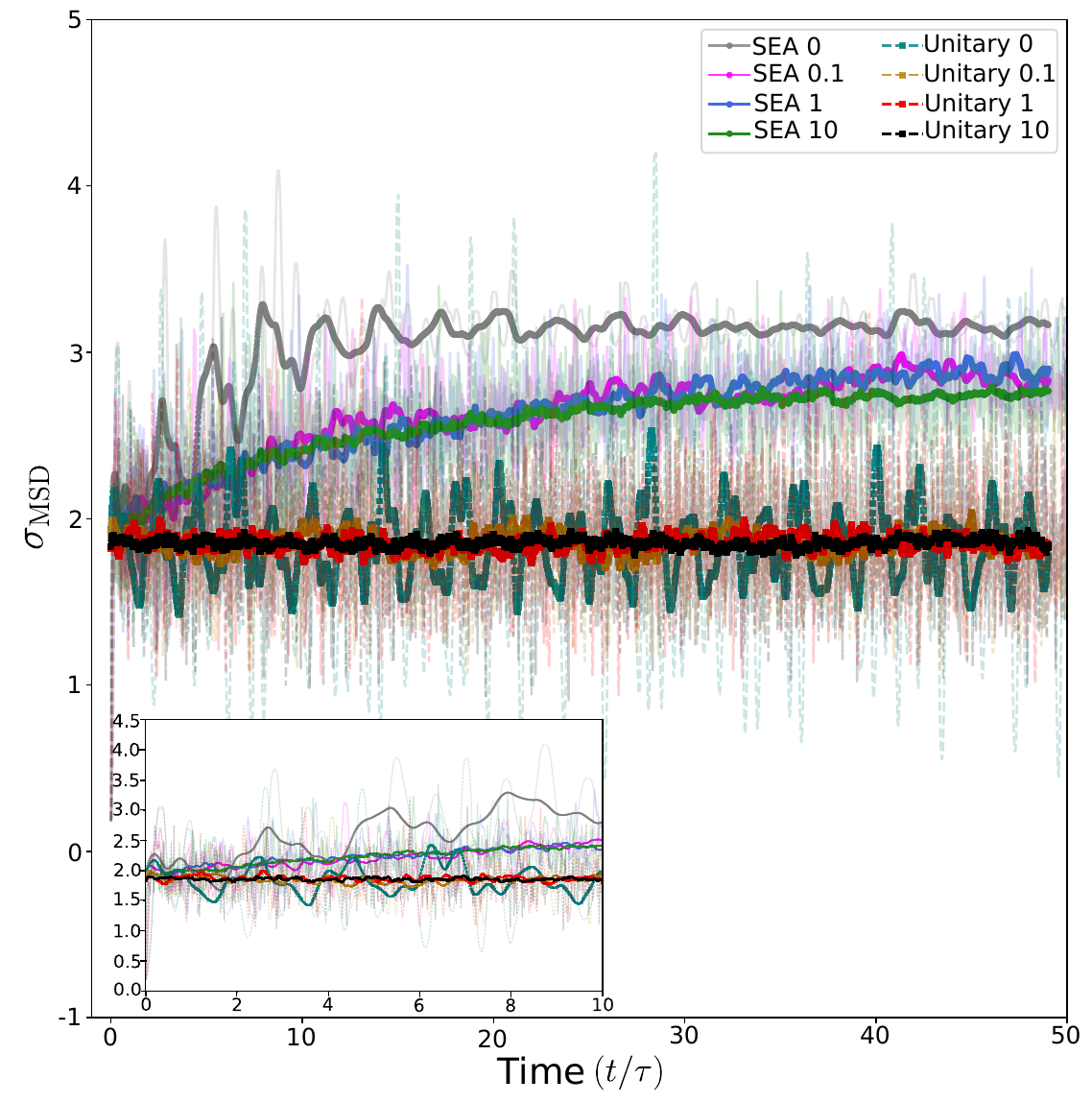}}\\ 
 \subfloat[ \label{fig:msd_HI} HI]{\includegraphics[width=\columnwidth]{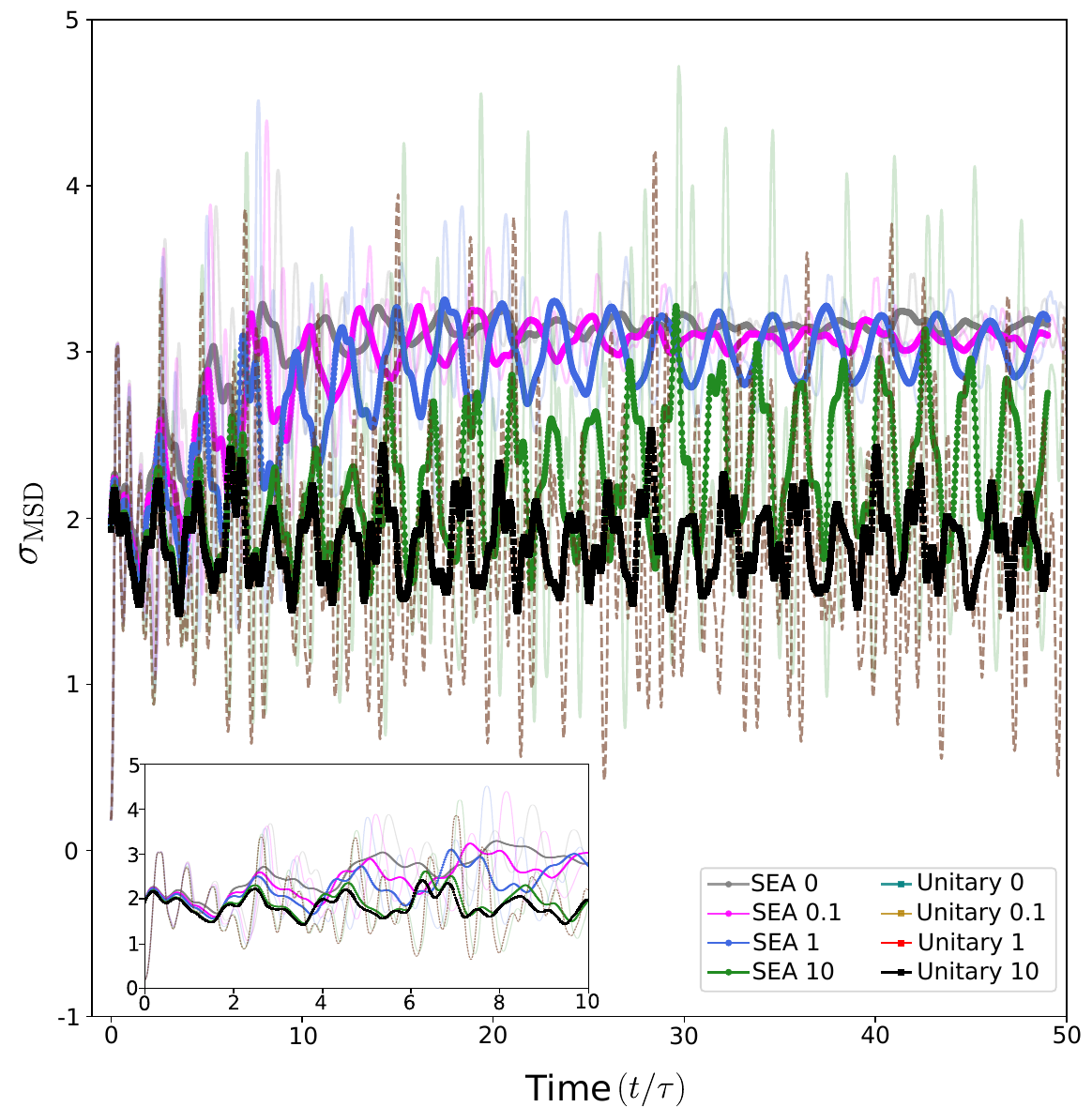}} 

\caption{Mean squared displacement ($\sigma_\text{MSD}$) computed from the two-walker JPD using Eq. (\ref{eq:msd}). Bold lines with markers represent the moving average, while transparent lines show the MSD evolution. (a) corresponds to the FI regime, and (b) to the HI regime. The numbers following SEA and Unitary in the legend indicate the values of $\alpha_i$'s, representing interaction strength. The insets show the early-time evolution of MSD.}
\label{fig:msd}
\end{figure}
This evolution of JPD is reflected in the time evolution of the marginal probability of the walkers. We show the marginal probability of the walkers in Fig. \ref{fig:marg_0_full}. As expected, the early time marginal evolution under unitary (panel (a) of Fig. \ref{fig:marg_0_full}) shows similar oscillations to the marginal evolution under SEA (panel (e) of Fig. \ref{fig:marg_0_full}). The late time marginal evolutions of unitary (panel (b) of Fig. \ref{fig:marg_0_full}) and SEA (panel (f) of Fig. \ref{fig:marg_0_full}) don't agree with each other, the unitary spikes are sharper in comparison (see the associated color bars for the difference in values). The interference-like patterns are also less in the SEA case. The smearing of peaks feature as seen in the JPD of Fig. \ref{fig:jpd_0_full} is also reflected in the marginals as seen in panels (c) and (d) of Fig. \ref{fig:marg_0_full}. The panels (d) are more smeared than the panels (c) in Fig. \ref{fig:marg_0_full}. In this case, we see in comparison to no interaction case (Fig. \ref{fig:marg_0_full}), under FI (Fig. \ref{fig:marg_10_full}), the marginal probability spreads faster. We observe this by noting how early the initial probability spike travels to the far end of the lattice. This is because the walkers are repulsively interacting with each other, and this interaction is causing the walkers to spread. We can also see how, in both cases, the marginal distribution is always peaked around the center (where the walk originated from) no matter how faint. This suggests that although the JPD shows rapid spread to the boundary, the marginal retains a dull peak at the origin of the walk. This is not only the case with FI, but also in the other cases as considered in the Table \ref{table:cases}. This is mostly true for the unitary case. In the case of SEA evolution, the probability of finding the walkers at the origin is less than the corresponding unitary scenarios, as it is more spread out because of the dissipation. \\
\begin{figure}[t!]
    \centering
    \includegraphics[width=\columnwidth]{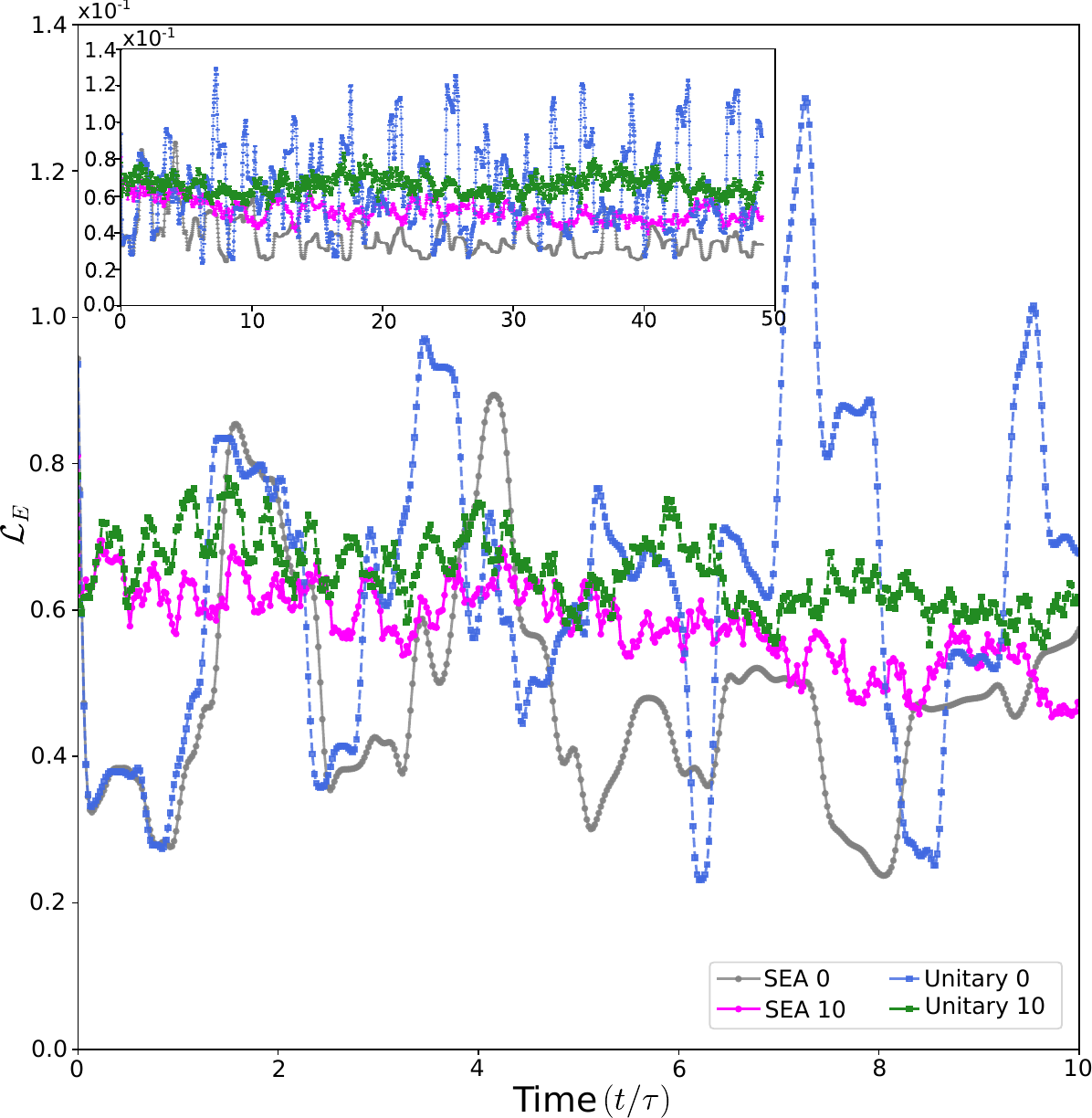}
    \caption{\label{fig:LE_full} Evolution of the moving average of Loschmidt echo over time (short timescale). In the legend, `0’ denotes no interaction, while `10’ indicates $\alpha_i=10$ in Eq. (\ref{eq:gen_total_ham}). This corresponds to the full interaction case (Table \ref{table:cases}). The inset shows the longer-time evolution.}
\end{figure}
Apart from various measures of probabilistic evolutions (\emph{e.g.}, JPD, marginals), we can characterize the walk by studying the mean square displacement (MSD) of the walks. This tells us the walkers' mean spread in space from their initial position. We can compute the MSD via the following equation ($m,n$ are the site numbers of the walkers):
\begin{equation}\label{eq:msd}
\sigma_{\text{MSD}} = \dfrac{1}{N}\sum_{m,n}^{N} (m-n)^2  \mathcal{P}^t_a(m,n)\,.
\end{equation}
If we consider the full interaction \emph{i.e.,} FI regime, we can see in Fig. \ref{fig:msd_FI} that the MSD under unitary evolution fluctuates less with increasing interaction strength and remains almost constant at all times of the evolution. On the other hand, the SEA-induced evolution becomes closer to unitary evolution as the interaction strength increases. Also, at FI, SEA MSD is higher than HI Fig. \ref{fig:msd}.\\
Besides MSD, we can also compute the Loschmidt echo (LE) of the walk, which measures the overlap between the initial ($\rho^0$) and time-evolved ($\rho^t$) density matrices, given by 
\begin{equation}\label{eq:lochsmidt}
  \mathcal{L}_E = \tr(\rho^0\rho^t).
\end{equation}
$\mathcal{L}_E$, tracks correlations between different times in the evolution. An alternative approach would be to compute LE using the trace of $\rho^{-t}\rho^t$, but in this work, we adopt the definition in Eq. (\ref{eq:lochsmidt}). For pure states, \(\mathcal{L}_E = 1\). If it remains at 1 (for pure initial states) or stabilizes at a value \(<\,1\) (for mixed states), the evolution is reversible. A gradual decrease in \(\mathcal{L}_E\) signals increasing irreversibility. Consider the expression,
\begin{figure}[t!]
    \centering
    \includegraphics[width=\columnwidth]{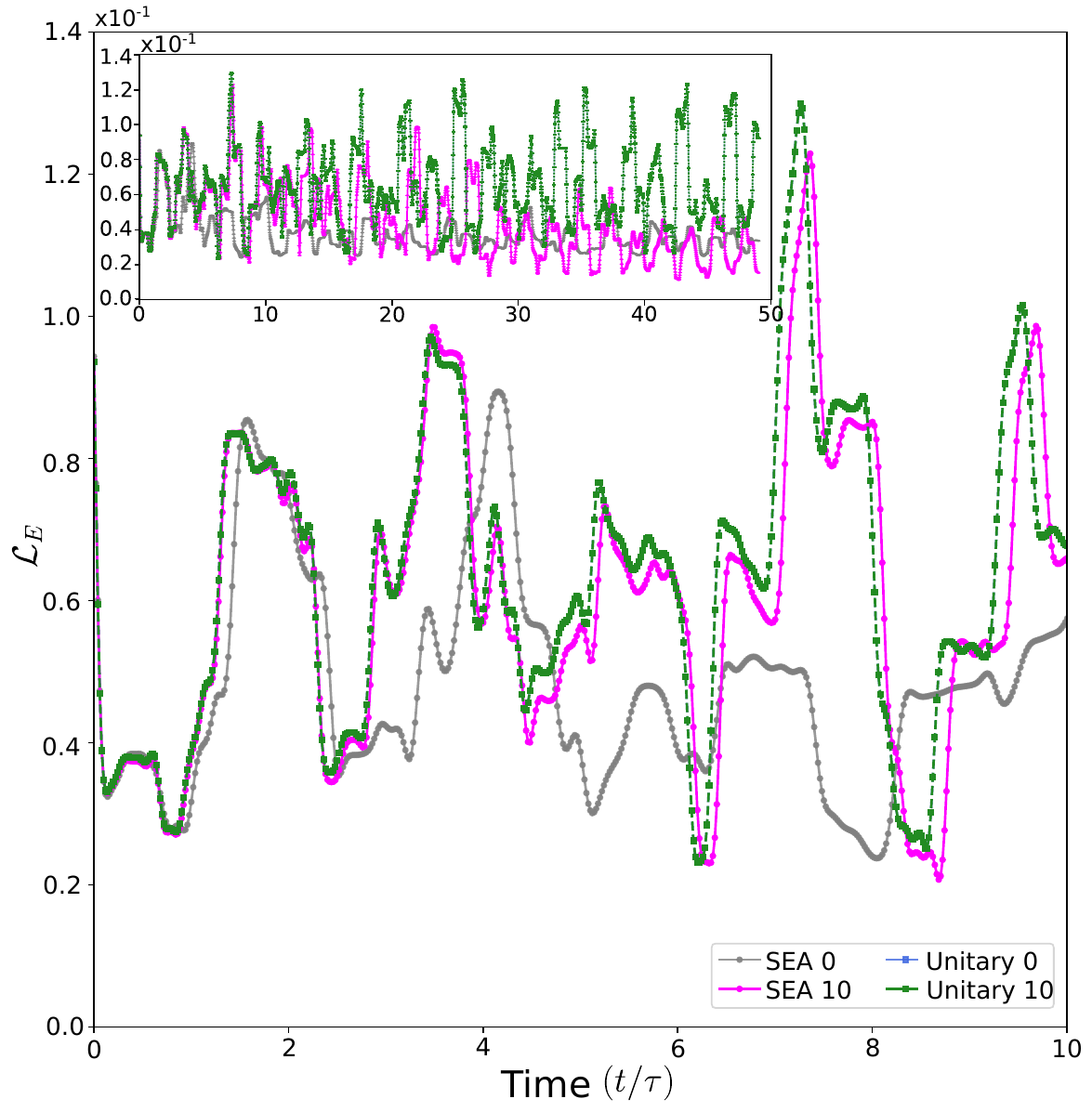}
    \caption{\label{fig:LE_hubbard} Evolution of the moving average of Loschmidt echo over time (short timescale). In the legend, `0’ denotes no interaction, while `10’ indicates $\alpha_1=10$ in Eq. (\ref{eq:gen_total_ham}). This corresponds to the Hubbard interaction case (Table \ref{table:cases}). The inset shows the longer-time evolution. }
\end{figure}
\begin{equation}\label{eq:le_expansion}
  \mathcal{L}_E = \tr(\rho^0\rho^t) = \tr(\rho^0\mathcal{U}_t\rho^0\mathcal{U}_t^\dagger).
\end{equation}
We used Eq. (\ref{eq:unitary_one_walker}) in the second equality.  Under unitary evolution with a pure initial state $\rho^0$, we have $\mathcal{L}_E = 1$.  However, in this work, we start with a mixed state, and since unitary evolution preserves purity, $\mathcal{L}_E$ remains constant but \(<\,1\).  If integrability is broken by adding interaction terms, $\mathcal{L}_E$ may decrease over time \cite{chenu_2018_quantum,serbyn_2017_loschmidt}. Let us look at Fig. \ref{fig:LE_full}. As time increases, $\mathcal{L}_E$ decreases for SEA evolution. It decreases faster in case of no interaction (inset of Fig. \ref{fig:LE_full}), and it decreases slower in case of full interaction. We also notice that $\mathcal{L}_E$ for unitary changes much more slowly and is usually greater than the SEA values. We also note that in the presence of interaction, the $\mathcal{L}_E$ for SEA is closer to that of the unitary value. If we consider the HI regime, we know that the non-SEA evolution is integrable, which can be seen from the $\mathcal{L}_E$ plot, as it remains unchanged upon varying interaction strength. We plot this in Fig. \ref{fig:LE_hubbard}. We see that the $\mathcal{L}_E$ varies differently in consideration to FI picture (Fig. \ref{fig:LE_full}). Also, the absence of extra interaction terms makes the unitary $\mathcal{L}_E$ much closer to the SEA $\mathcal{L}_E$ in the HI regime at initial times. However, as time progresses, SEA evolution becomes more dissipated with interaction than even the free-from-interaction case in HI regime (see inset of Fig. \ref{fig:LE_hubbard}). \\
So far, we have seen how the characteristic measures of the walk differ under the influence of unitary and SEA dynamics. Now, we focus our attention on entropy. The principal tenet of SEA being the generation of entropy, we expect to see entropy gain and the corresponding decrease in mutual information ($\mathcal{M}(\rho)$) defined as ($\Boltz=1$)
\begin{equation}\label{eq:mi}
\mathcal{M}(\rho) = \Boltz\tr(\rho\ln(\rho))-\Boltz\sum_{\J=1}^{M}\tr(\rho_\J\ln(\rho_\J)).
\end{equation}
\begin{figure}[t!]
\centering
\includegraphics[width=\columnwidth]{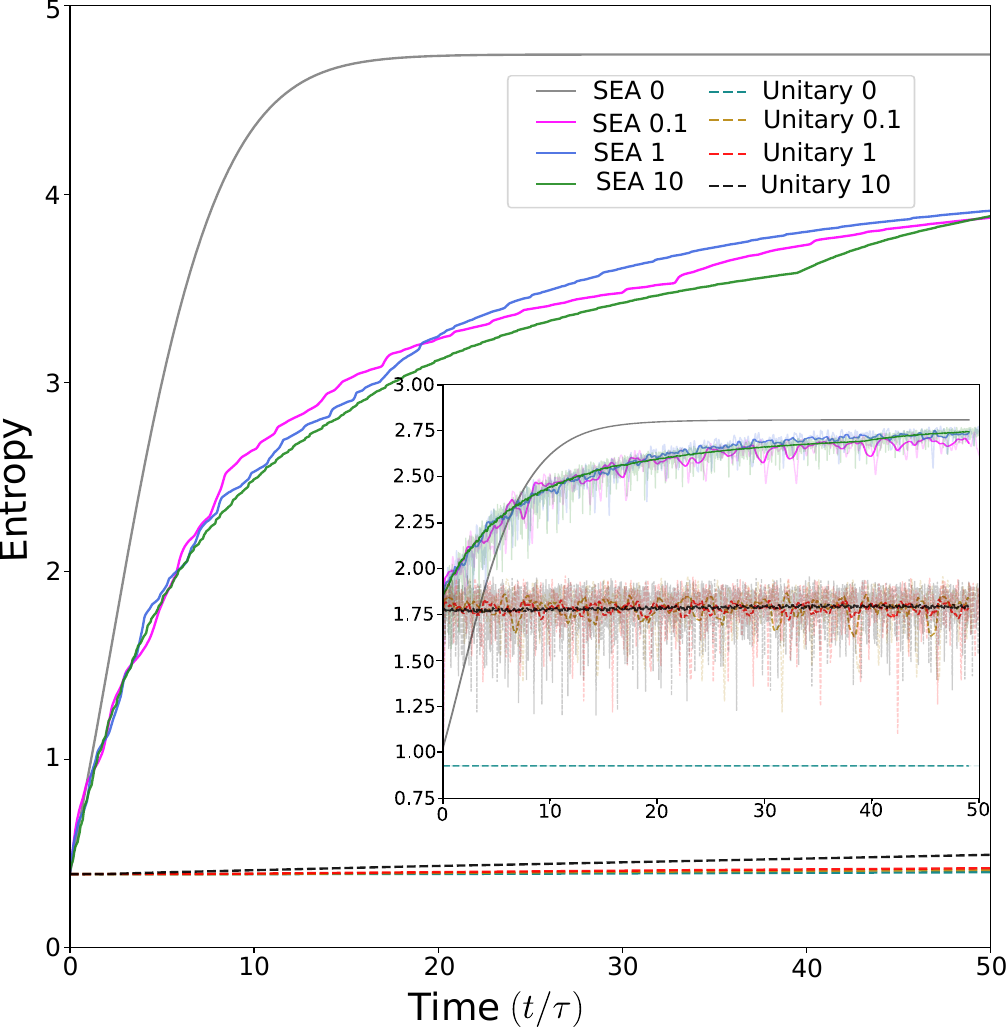}
\caption{\label{fig:entropy_FI}  Evolution of entropy in the two-walker system under the FI regime. The values of \(\alpha_i\) are indicated in the legend next to the SEA and Unitary labels. The inset shows the entropy evolution of subsystem A, where the original entropy values are plotted with higher transparency, while the moving average is highlighted in bold.}
\end{figure}
The decrease in $\mathcal{M}(\rho)$ is a measure of the loss of correlation between the walkers. Especially under SEA evolution, since no extra correlation is being created, this should be the case. We can see this in Fig. \ref{fig:entropy_FI}, where we plot the time evolution of the entropy of the two-walker system under both unitary and SEA in the FI regime. As interaction strength increases, the unitary evolution slightly departs from integrability, as seen from the deviation of otherwise constant entropy in Fig. \ref{fig:entropy_FI}. What happens in other interaction regimes? For instance, in the FIFH regime, where the weights of correlated hopping and on-site potentials are tuned relative to the hopping terms---a variation of the FI regime---we observe in Fig. \ref{fig:entropy_FIFH} that entropy growth is influenced by interaction strength. In unitary evolution, entropy increases as interaction terms break integrability, leading to deviations from integrable dynamics. However, this is not the case for SEA. Without interaction, SEA exhibits faster entropy production, but as interaction strength increases, entropy production slows down, delaying the onset of thermalization. \\
\begin{figure}[t!]
  \centering
  \includegraphics[width=\columnwidth]{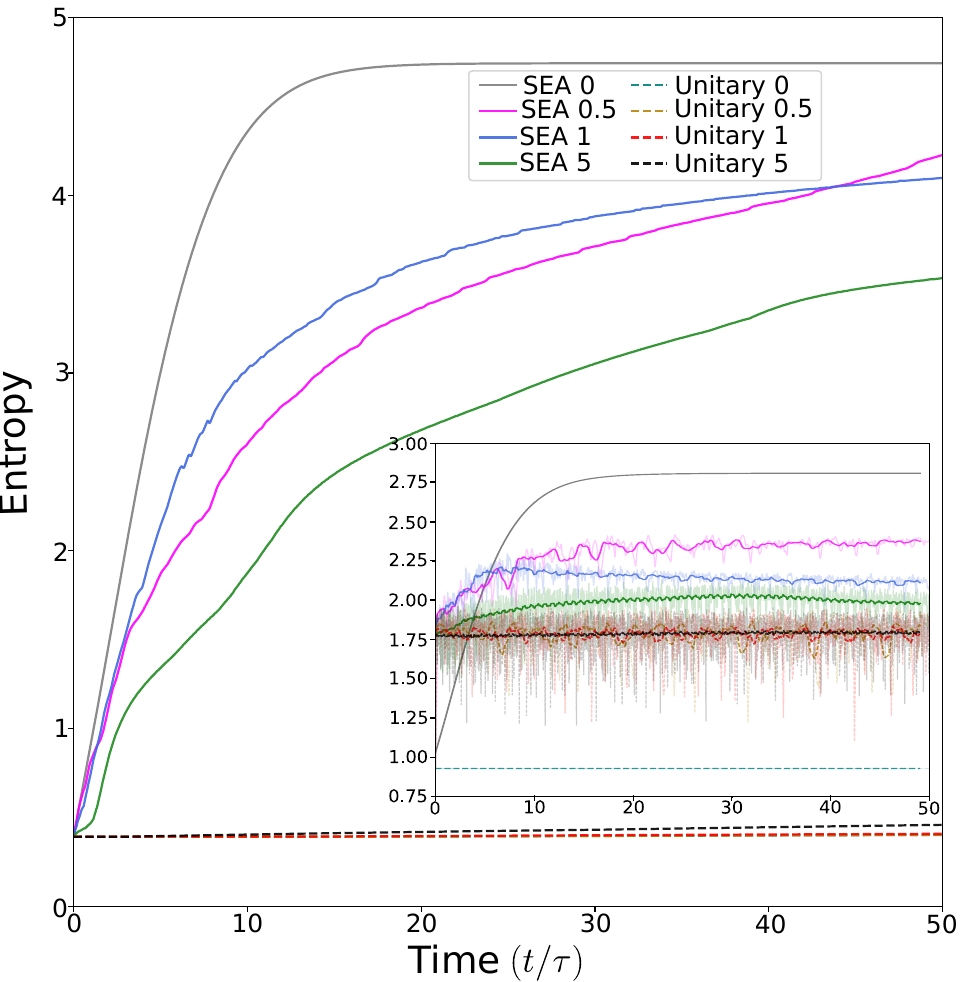}
  \caption{\label{fig:entropy_FIFH}   Evolution of entropy in the two-walker system under the FIFH regime. The values of \(\alpha_1\) and \(\alpha_4\) are indicated in the legend next to the SEA and Unitary labels, with \(\alpha_2=\alpha_3=0.1\) fixed. The inset shows the entropy evolution of subsystem A, where the original entropy values are plotted with higher transparency, and the moving average is highlighted in bold.}
\end{figure}
We can infer when the system reaches thermalization by examining the evolution of $\mathcal{M}(\rho)$. Once it saturates, no further correlations are lost, signaling the onset of thermalization. Fig. \ref{fig:MI} reveals that even a small interaction induces significant correlation buildup. As interaction strength increases, $\mathcal{M}(\rho)$ saturates later, particularly in Fig. \ref{fig:MI_FI}, where all $\alpha_i$'s have equal weight. However, when the weight ratio is skewed, correlation loss becomes extensive---surpassing even the no-interaction case, as seen in Fig. \ref{fig:MI_FIFH}. Notably, while $\mathcal{M}(\rho)$ has not yet saturated within the timescale considered, its eventual saturation value appears significantly lower in the FIFH case than in FI. This suggests a direct dependence of $\mathcal{M}(\rho)$'s saturation value on the relative strengths of different interaction terms.
\begin{figure}[t!]
  \centering
 \subfloat[ \label{fig:MI_FI} FI]{\includegraphics[width=\columnwidth]{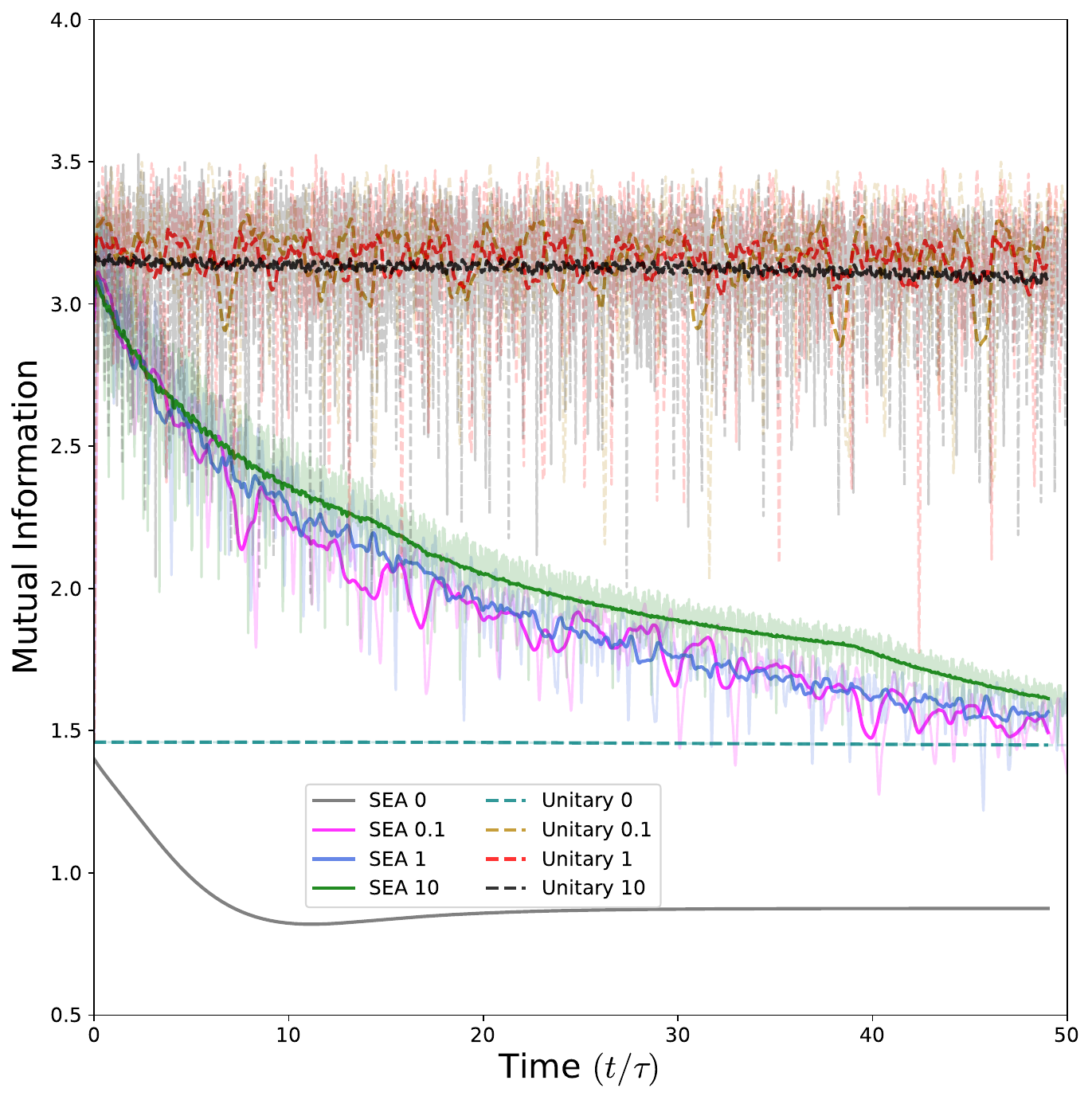}}\\ 
 \subfloat[ \label{fig:MI_FIFH} FIFH]{\includegraphics[width=\columnwidth]{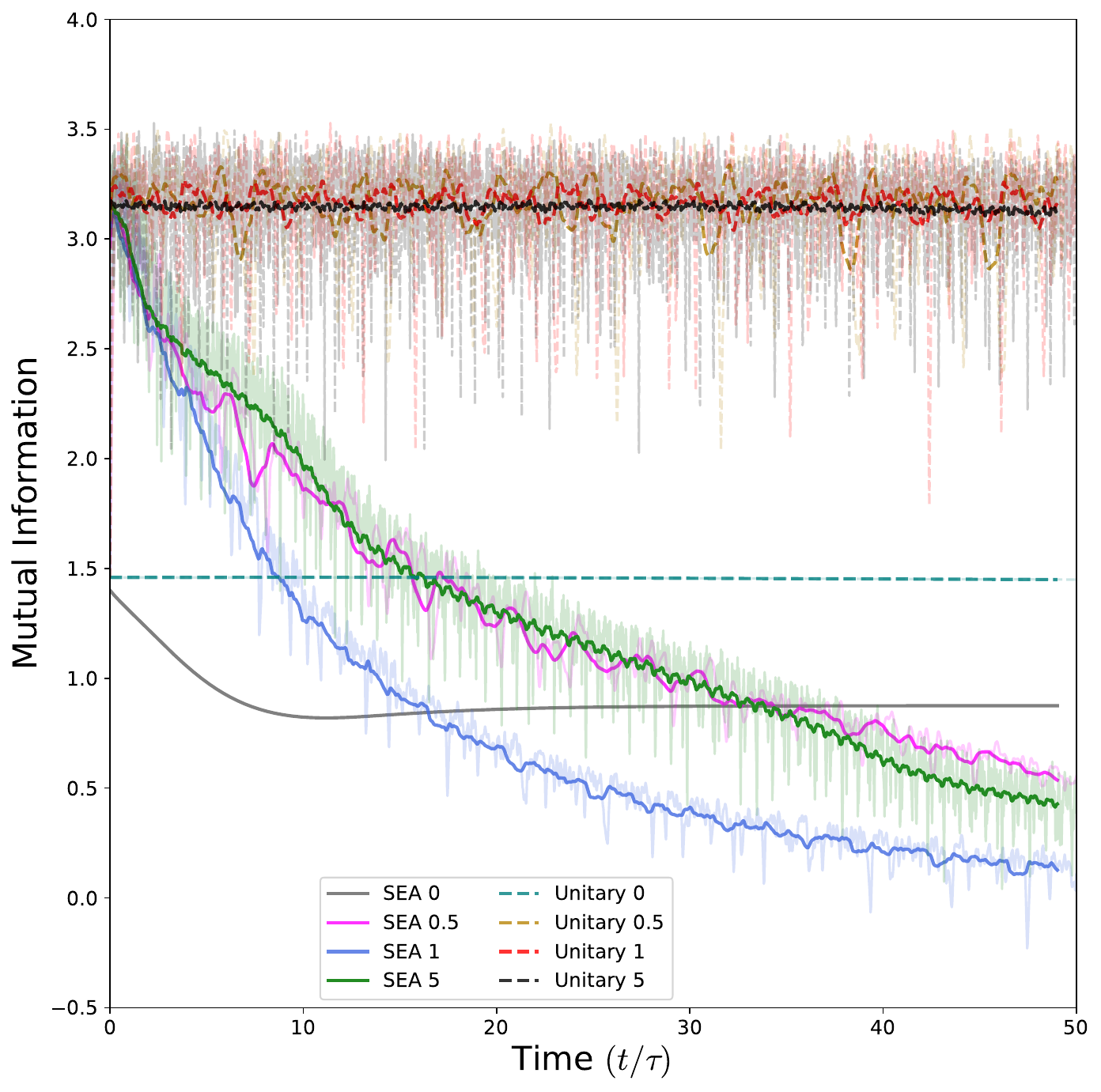}} 
  \caption{Mutual information ($\mathcal{M}(\rho)$) computed using Eq. (\ref{eq:mi}). Bold lines represent the moving average, while transparent lines show the evolution of $\mathcal{M}(\rho)$. (a)  corresponds to the FI interaction regime, and (b) to the FIFH regime. In (a), the numbers following SEA and Unitary in the legend indicate the values of $\alpha_i$'s representing interaction strength. In (b), they denote $\alpha_1 = \alpha_4$, with $\alpha_2=\alpha_3=0.1$.}
  \label{fig:MI}
\end{figure}

\section{\label{sec:discussions} Discussion and Conclusion}

In this work, we theoretically explored the evolution of two walkers under dissipative steepest entropy ascent (SEA) dynamics and compared the results with bare unitary evolution. By examining various interaction regimes, we analyzed key characteristics of the walk, including the joint probability distribution (JPD), mean squared displacement ($\sigma_\text{MSD}$), and the Loschmidt echo ($\mathcal{L}_E$). We also investigated entropy changes under unitary and SEA evolutions and examined mutual information ($\mathcal{M}(\rho)$) to gain insight into the system's thermalization.\\
We began by introducing the Beretta steepest entropy ascent equation of motion (BSEA EoM) \cite{beretta_2020_fourth} for two components of a composite system. However, our work does not involve an interacting reservoir for thermalization. Instead, we quench the system from a pure to a mixed state by introducing noise (see Eq. (\ref{eq:initial_rho})),without detailing the quenching process. Post-quench, we assume that the system evolves in isolation. Unlike the Lindblad master equation, our approach does not involve continuous interaction with an external environment, eliminating the need for the Markovian approximation. In the SEA framework, the system is initialized in a mixed state, which means that it is already out of equilibrium. As a result, it undergoes further dissipation, following the SEA principle of maximum local entropy generation while respecting local conservation laws (e.g., probability, energy). \\
Through the evolution of JPD and the corresponding marginal probability distribution, we demonstrate that SEA evolution results in a greater probability spread across the lattice. While interaction strengths and regimes influence this spread, the overall trend remains consistent. The extent of probability distribution spreading depends on the interaction regime, as shown in Fig. \ref{fig:jpd_10_full}. We examine the marginal probability distribution at both early and late times, as depicted in Figs. \ref{fig:marg_0_full} and \ref{fig:marg_10_full}. In SEA evolution, we observe smearing in late-time marginals, indicating increased mixing. This probability spread follows the same patterns observed in JPD evolution.\\
To better understand the nature of the walk, we examine the evolution of $\sigma_\text{MSD}$, which quantifies the overall spread of the walkers and the system’s behavior. Under unitary evolution, $\sigma_\text{MSD}$ remains nearly constant across all regimes, with minor fluctuations. In contrast, SEA evolution exhibits a significant spread, consistent with JPD and marginal results, and is influenced by interaction types. For instance, when the on-site potential dominates, SEA and unitary spreads converge at higher \(\alpha_1\) values (Fig. \ref{fig:msd_HI}). Increasing the weights of on-site and correlated hopping terms \(\alpha_1,\, \alpha_4\) accelerates walker spread in the CHI regime. The full interaction regime exhibits a high MSD (Fig. \ref{fig:msd_FI}), which can be further adjusted in the FIFH regime, where on-site and correlated hopping terms dominate. Across all interaction regimes, SEA spread \(\sigma_\text{MSD}\) increases in the order: FIFH $<$ FI $<$ CHI $<$ HI. \\
Regarding the Loschmidt echo, we see that the SEA evolution leads to a significant loss of coherence, which keeps increasing with time. This starkly contrasts with the unitary evolution, where coherence is preserved for weaker interactions. As more interaction terms are introduced, we see a steady decrease in $\mathcal{L}_E$ of both evolutions (see Fig. \ref{fig:LE_full} for FI and Fig. \ref{fig:LE_hubbard} for the HI regimes). Numerical results show that in different regimes, \(\mathcal{L}_E\) under SEA compared to bare unitary decreases in the order: HI $>$ CHI $>$ FIFH $>$ FI (largest to smallest deviation at late times). This suggests that as interaction terms increase, making unitary evolution more non-integrable, SEA \(\mathcal{L}_E\) approaches unitary \(\mathcal{L}_E\) over time. The ordering of \(\sigma_\text{MSD}\) and \(\mathcal{L}_E\) confirms that SEA evolution is more sensitive to interaction terms than unitary evolution.\\
Furthermore, our analysis of entropy and mutual information $\mathcal{M}(\rho)$ supports the conclusions above. The entropy plots clearly illustrate how the onset of entropy saturation (maximum achievable entropy) depends on interaction strength. Within the time scales and interaction regimes considered, the free SEA entropy reaches its maximum. This trend is also evident in the entropy of the subsystem, as shown in the insets of Figs. \ref{fig:entropy_FI} and \ref{fig:entropy_FIFH}. Interestingly, we confirm that subsystem entropy not only decreases with an increasing $\alpha_4/\alpha_2$ ratio but also saturates later, indicating delayed thermalization. A similar analysis of $\mathcal{M}(\rho)$ provides further insight into the thermalization process. The observed decrease in subsystem entropy (and overall entropy in Fig. \ref{fig:entropy_FIFH}) aligns with the declining trend of mutual information. Notably, mutual information does not reach saturation within the time scale considered, suggesting that longer time-scale studies are needed to fully understand the thermalization process.\\
In this work, we aim to extend the SEA formalism to discuss the evolution of multi-walker quantum systems. In performing so, we investigated the varied interaction regimes and their effects on the characteristic measures of the walk. We have also studied how the entropy saturation is set and how the introduction of the interaction delays the same onset, as hinted in \cite{ray_2025_nosignalinga}. We can implement two walker continuous-time set-up under SEA on superconducting qubit platforms such as the one described in Ref.~\cite{yan_2019_Stronglycorrelated}. A theoretical formalism using this set-up can be established from the work of Ref.~\cite{montanez-barrera_2022_decoherence}. Our results motivates us to continue our research to perform a detailed analysis on the various time of thermalization for various interaction strengths under SEA, and how different those thermalization times are from the unitary case. We will also be interested in knowing if the scaling of thermalization times with system sizes is the same for both unitary and SEA evolutions. And that will lead us to understand if there exist universal scaling laws in the SEA evolution thus considered. We can also include more than two walkers using the concept of hypoequilibrium~\cite{li_2016_steepestentropyascent} to solve the complicated many-body SEA equation and simulate dissipative continuous-time quantum walk in this regime.

\section*{\label{Acknowledge}Acknowledgment}

RKR acknowledges financial support from the Institute for Basic Science (IBS) in the Republic of Korea through Project No. IBS-R024-D1. RS acknowledges the financial support from the Indian Science \& Engineering Research Board (SERB) grant CRG/2022/008345.

\appendix
\section{\label{app:two_component}Derivation of two-component equation of motion}
Here we derive the expression for the dissipation operator, described in Eq.~\eqref{eq:DJ_compact}, for the dynamics of the composite system. The detailed derivation of the dissipation operator for the single particle case is described in the Appendices of Ref.~\cite{ray_2022_steepestb}. Here, in this Appendix, to avoid repetition, we highlight the core tenets of the steepest entropy ascent formalism. Thereafter, we discuss how we follow the SEA formalism to arrive at the composite equation of motion. 

\subsection{The SEA principle}
The idea of SEA is embedded in variational principle. We seek to identify the path in a given state space (defined by density matrices) that will satisfy the following properties:
\begin{enumerate}
    \item The conserved quantities that generate the motion (energy, momentum, number of particles, probability, etc.) are invariant under time evolution.
    \item The local entropy production rate is always positive semidefinite.
    \item The evolution proceeds in the direction of maximum entropy production.
\end{enumerate}
Keeping these in mind, we note that the set of conserved quantities defining the constraints of the evolution lie in a hyperplane. The entropy functional generally does not lie in the normal direction to this hyperplane. Therefore, to find the gradient of the entropy functional projected relative to these constraints, we seek the direction of the projected normal vector, which defines the SEA direction. To perform this, we apply calculus of variations to find the suitable Lagrange's multipliers and then use that to write the equation of motion. To this extent, we proceed as under.

\subsection{The composite dissipation structure}
The Steepest Entropy Ascent (SEA) equation for composite systems introduces a structure-dependent dissipation term (as in Eq.~\eqref{eq:comp_sea_simplified}):
\begin{equation}\label{appeq:sea_compact}
    \dv{\rho}{t} = -\imi[H, \rho] - \sum_{J=1}^{M} \{ \mathcal{D}_\J, \rho_\J \} \otimes \rho_\Jbar
\end{equation}
where $\mathcal{D}_\J$ is the local dissipation operator on the subsystem $J$, and $\rho_\J = \Tr_{\Jbar}(\rho)$ is the reduced density operator of $\J$. As is important in the context of SEA, the rate of change of the overall system entropy $s(\rho) = -k_B \Tr(\rho \ln \rho)$ is given by~\cite{beretta_1985_quantuma, beretta_2010_maximuma, ray_2025_nosignalinga}:
\begin{equation}\label{appeq:SEA_entropy_prod}
    \frac{ds(\rho)}{dt} = - \sum_\J \Tr\left[ \{ \mathcal{D}_\J, \rho_\J \} (S)^J_\rho \right].
\end{equation}
Now, if we trace the subsystem $\Jbar$ in $\Hil_\Jbar$ in Eq.~\eqref{eq:comp_sea_simplified}, or Eq.~\eqref{appeq:sea_compact} above, we get the corresponding local evolution for the subsystem $\J$:
\begin{equation}
    \frac{d\rho_\J}{dt} = -\imi[H_\J, \rho_\J] - \imi \Tr_{\Jbar}([V, \rho]) - \{ \mathcal{D}_\J, \rho_\J \}
\end{equation}
So far, this is a general dissipative dynamical equation. The crucial SEA assumption will be implemented below.

\subsection{The SEA variational principle}
To maintain a positive semidefinite nature of the density matrices during the evolution, we define the generalized square root:
\begin{equation}
    \gamma_\J = \sqrt{\rho_\J} U\,,
\end{equation}
where $U$ is an arbitrary unit operator. It implies $\rho_\J = \gamma_\J \gamma_\J^\dagger$. Using this decomposition of $\rho$, we can expand the dissipator anti-commutator as
\begin{equation}\label{appeq:SEA_Composite_gamma}
    - \{ \mathcal{D}_\J, \rho_\J \} = \dot{\gamma}_\J^d \gamma_\J^\dagger + \gamma_\J \dot{\gamma}_\J^{d\dagger}, \text{ with } \dot{\gamma}_\J^d = - \mathcal{D}_\J \gamma_\J
\end{equation}
We also define the symmetric inner product $( \cdot | \cdot )$ in the set $\mathcal{L}(\Hil_\J)$ of linear operators on $\Hil_\J$ as~\cite{beretta_2014_steepesta, ray_2022_steepestb}
\begin{equation}\label{appeq:innerProduct}
	( X | Y)=\half\Tr (X^{\dagger} Y + Y^{\dagger} X ).
\end{equation}
We define the inner product so that the unit trace condition for $\rho_\J$ is rewritten as $( \gamma_\J | \gamma_\J)=1$. This ensures that the trajectories\footnote{In the SEA evolution, since we are invested in `paths', `maximization', and similar distance-related concepts, we need to define the distance in this context.} traced by the evolution of $ \gamma_\J$ are confined to the unit sphere in $\mathcal{L}(\Hil_\J)$ \footnote{The Bloch-sphere in case of a single qubit.}. We are now in a position to define the distance traveled between $t$ and $t{\,+\,}\d t$ along these trajectories as ---
\begin{equation}\label{appeq:distance}
	\d\ell_\J = \sqrt{(\dot\gamma_\J|\,\hat G_\J(\gamma_\J)\,|\dot\gamma_\J)}\, \d t\,,
\end{equation}
where $\hat G_\J(\gamma_\J)$ is some real, dimensionless, symmetric and positive definite operator on $\mathcal{L}(\Hil_\J)$ (superoperator on $\Hil$) that plays the role of a local metric tensor field (and may be a non-linear function of $\gamma_\J$)~\cite{beretta_1985_quantuma, beretta_2010_maximuma, beretta_2014_steepesta, ray_2022_steepestb, ray_2025_nosignalinga}. The rate of change of the overall entropy of the system, $s(\rho)$, Eq.~\ref{appeq:SEA_entropy_prod}, and of the mean value of the overall system of conserved properties $c_k(\rho)=\Tr(\rho C_k)$, where $\comm*{C_k}{H}=0$, can be rewritten as
\begin{align}
	\dv{s(\rho)}{t}&= \sum_{\J=1}^M\dot{s}|_\J \qquad \dot{s}|_\J=\Big(2(S)^\J_\rho\gamma_\J\Big|\dot\gamma^{d}_\J\Big)\,, \label{appeq:SEA_entropy_prod_gamma}\\
	\dv{c_k(\rho)}{t}&= \sum_{\J=1}^M\dot{c_k}|_\J\qquad \dot{c_k}|_\J=\Big(2(C_k)^\J_\rho\gamma_\J\Big|\dot\gamma^{d}_\J\Big)\,, \label{appeq:constants_gamma}
\end{align}
exhibiting additive contributions from the subsystems. Finally, we implement the SEA principle. According to this, the time evolution ensures that the ``direction of change'' of the local trajectory $\gamma_\J(t)$, influenced by the dissipative part of the dynamics, maximizes the local contribution $\dot{s}|_\J$ to the overall system's entropy production rate. This happens under the condition that the constraints of the motion $\dot{c_k}|_\J=0$ have no local contribution to the rate of change of the global constants of the motion. We state the variational principle that yields expressions for $\dot\gamma^{d}_\J$'s and the $\mathcal{D}^\J_\rho$,  which define the composite-system version of the SEA equation of motion, as follows: 
\begin{equation}
   \max_{\dot\gamma^{d}_\J}\ \Upsilon_\J = \dot{s}|_\J - \sum_k \beta_k^\J \dot{c}_k|_\J - \frac{k_B \tau_\J}{2} ( \dot{\gamma}_\J^d | \hat{G}_\J | \dot{\gamma}_\J^d ),
\end{equation}
Here $\beta_k^\J$ is the $k^\Th$ Lagrange multiplier associated with $k^\Th$ the conserved quantity $c_k(\rho)$ for the subsystem $\J$. And $\tau_\J$ is the same associated with the relaxation time of the subsystem $\J$~\cite{beretta_2014_steepesta, ray_2022_steepestb}. Solving the variational problem yields the following.
\begin{equation}\label{appeq:SEA_general_supermetric}
    |\dot{\gamma}_\J^d) = \frac{1}{k_B \tau_\J} \hat{G}_\J^{-1} | 2 (M)^J_\rho \gamma_\J )
\end{equation}
And the locally perceived non-equilibrium Massieu operator~\cite{beretta_1985_quantuma}:
\begin{equation}
    (M)^J_\rho = (S)^J_\rho - \sum_k \beta_k^J (C_k)^J_\rho
\end{equation}
The Lagrange multipliers $\beta_k^\J$ (implicit in $(M)^\J_\rho$) are the solution of the system of equations, obtained by substituting Eq.~(\ref{appeq:SEA_general_supermetric}) into the conservation constraints,
\begin{equation}\label{appeq:multiplierssuper} 
	\Big((C_\ell)^\J_\rho\gamma_\J\Big|\hat G_\J^{-1}\Big|(M)^\J_\rho\gamma_\J\Big)=0\quad \forall \ell\,.
\end{equation}
We can use Cramer's rule to solve this system of equations for the $\beta_k^\J$'s. As seen in Eqs.~\eqref{eq:app1_betaJ1}-\eqref{eq:app1_betaJ2}, the $\beta_k^\J$'s are nonlinear functionals of $\rho$ that may be interpreted as ``local non-equilibrium entropic potentials'' conjugated with the conserved properties. For example, for $C_2=H$, the conservation of energy, $\beta_2^\J$, plays the role of ``local non-equilibrium inverse temperature'' conjugated with the locally perceived energy, and for the stable equilibrium states of the SEA dynamics, it coincides with the thermodynamic inverse temperature $\Boltz\beta_\J$.

Finally, Eq.~\eqref{appeq:sea_compact}, the equation of motion results independent of the arbitrary unitary operators $U$ used in the definition of $\gamma_\J$ we can restrict the choice of the metric superoperator $\hat G_\J$. We assume that $\hat G_\J= L_\J^{-1} \hat I_\J$ with $L_\J$ some strictly positive Hermitian operator on $\Hil_\J$, possibly a nonlinear function of $\rho_\J$, so that $\hat G_\J|X) = |L_\J^{-1} X)$, $\hat G^{-1}_J|X) = |L_\J X)$, 
\begin{equation} 
	(X\gamma_\J|\hat G^{-1}_J|Y\gamma_\J) = \half\Tr\left[ \rho_\J(X^\dagger L_\J Y+Y^\dagger L_\J X)\right].
\end{equation}
Recalling Eq.~\eqref{appeq:SEA_Composite_gamma} with Eq~\eqref{appeq:SEA_general_supermetric} , the dissipative term in Eq.~\eqref{appeq:sea_compact} becomes , 
\begin{equation}\label{appeq:SEA_general} 
	-\acomm*{\mathcal{D}^\J_\rho}{\rho_\J}=\frac{2}{\Boltz\tau_\J}\left[L_\J (M)^\J_\rho\rho_\J +\rho_\J (M)^\J_\rho L_\J\right]\,,
\end{equation}
and the system of equations that determines the Lagrange multipliers  $\beta_k^\J$ in $(M)^\J_\rho$ is
\begin{equation}\label{appeq:multipliers} 
\Tr\left(\rho_\J\Big[(C_\ell)^\J_\rho L_\J(M)^\J_\rho+(M)^\J_\rho L_\J(C_\ell)^\J_\rho\Big]\right)=0\quad \forall \ell.
\end{equation}
Therefore, all terms have become dependent on the local state operator $\rho_\J$. Now that we have the form of $\mathcal{D}_\J$, we now show how to obtain Eq.~\eqref{eq:DJ_compact} from these relations. We need to define our metric, and explicitly mention the conserved quantities.

\subsection{Simplest composite-system BSEA equation}

We make the following assumptions ---

\begin{enumerate}
    \item Uniform Fisher-Rao metric: $\hat{G}_\J = \hat{I}_\J \Rightarrow L_\J = I_\J$
    \item Conserved properties: $C_1 = I$, $C_2 = H$
\end{enumerate}

Under the first assumption, we get 
\begin{equation}\label{appeq:DJ_Def}
    \mathcal{D}^\J_\rho =-\frac{2}{\Boltz\tau_\J}\left[(S)^\J_\rho- {\textstyle \sum_k}\beta^\J_k(C_k)^\J_\rho\right].
\end{equation}
Using Eq.~\eqref{appeq:DJ_Def} with Eq.~\eqref{appeq:multipliers} along with $\Boltz =1$ we recover Eq.~\eqref{eq:app1_omegaJ}-\eqref{eq:app1_betaJ2}. The second assumption simply implies 
\begin{equation}
    \mathcal{D}_\J = -\frac{2}{\tau_\J} \left[ (S)^J_\rho - \beta_1^\J I_\J - \beta_2^\J (H)_\J \right].
\end{equation}

\bibliography{refer.bib}

\end{document}